\documentclass{article}
\usepackage{amssymb}
\usepackage{amsmath}
\usepackage{amscd}
\usepackage{amsthm}
\usepackage[latin9]{inputenc}

\newtheorem{theorem}{Theorem}
\newtheorem{definition}[theorem]{Definition}

\newtheorem{corollary}[theorem]{Corollary}
\newtheorem{proposition}[theorem]{Proposition}
\newtheorem{lemma}[theorem]{Lemma}

\newtheorem{question}[theorem]{Open Question}

\newcommand{\pls}{\textrm{PLS}}
\newcommand{\ppad}{\textrm{PPAD}}
\newcommand{\fixp}{\textrm{FIXP}}
\newcommand{\mto}{\rightrightarrows}
\newcommand{\dom}{\operatorname{dom}}
\newcommand{\Dom}{\operatorname{Dom}}

\newcommand{\codom}{\operatorname{CDom}}
\newcommand{\mult}{\textnormal{\emph{Mult}}}
\newcommand{\rel}{\textnormal{\emph{Rel}}}
\newcommand{\set}{\textnormal{\emph{Set}}}
\newcommand{\parti}{\textnormal{\emph{Par}}}
\newcommand{\id}{\textnormal{id}}

\newcommand{\Baire}{\mathbb{N}^\mathbb{N}}
\usepackage{tikz}
\usetikzlibrary{positioning}

\begin{document}

\title  {Many-one reductions and the category of multivalued functions}
\author{Arno Pauly\\University of Cambridge\\Arno.Pauly@cl.cam.ac.uk}
\date{}

\maketitle

\begin{abstract}
Multi-valued functions are common in computable analysis (built upon the Type 2 Theory of Effectivity), and have made an appearance in complexity theory under the moniker \emph{search problems} leading to complexity classes such as $\ppad$ and $\pls$ being studied. However, a systematic investigation of the resulting degree structures has only been initiated in the former situation so far (the Weihrauch-degrees).

A more general understanding is possible, if the category-theoretic properties of multi-valued functions are taken into account. In the present paper, the category-theoretic framework is established, and it is demonstrated that many-one degrees of multi-valued functions form a distributive lattice under very general conditions, regardless of the actual reducibility notions used (e.g.~Cook, Karp, Weihrauch).

Beyond this, an abundance of open questions arises. Some classic results for reductions between functions carry over to multi-valued functions, but others do not. The basic theme here again depends on category-theoretic differences between functions and multi-valued functions.
\end{abstract}

\tableofcontents

\section{Introduction}
\subsubsection*{What are multi-valued functions?}
A (partial) multi-valued function $f : \subseteq A \mto B$ is just a set $f \subseteq A \times B$ -- i.e.~a relation. However, the category of multi-valued functions is not the category of relations! We write $f(a)$ for $\{b \in B \mid (a, b) \in f\}$ and $\dom(f) = \{a \in A \mid \exists b \in f(a)\}$. Then the composition of multi-valued functions $f : \subseteq A \mto B$, $g : \subseteq B \mto C$ is defined via $c \in (g \circ f)(a)$ iff $f(a) \subseteq \dom(g)$ and $\exists b \in f(a)$ s.t.~$c \in g(a)$. In the usual definition of the composition for relations, the former condition is absent!

The intended interpretation of a multi-valued function $f : \subseteq A \mto B$ is that it links problem instances to solutions. This draws interest to the following partial order:
$$f \preceq g \ \Leftrightarrow \ \dom(f) \subseteq  \dom(g) \wedge g_{|\dom(f)} \subseteq f$$
We can read $f \preceq g$ as $f$ \emph{is easier as} $g$: There may be fewer instances for $f$ than for $g$, and a solution to a problem instance in $g$ is a solution for it in $f$, too, where applicable. This has the consequence that any procedure solving $g$ also solves $f$.

For any two multi-valued functions $f, g : \subseteq A \mto B$ there exists a hardest multi-valued function easier than both, i.e.~there are binary infima w.r.t.~$\preceq$. These are given by $f \wedge g = (f \cup g)_{|\dom(f) \cap \dom(g)}$.

\subsubsection*{Why use them?}
First, multi-valued functions are natural: From elimination orders on graphs over Nash equilibria in games to fixed points of continuous mappings, there are plenty of problems without a natural way to specify the desired solution uniquely. In fact, if one accepts their formulation as multi-valued functions, one can even prove that the latter two are non-equivalent to any function \cite{paulyincomputabilitynashequilibria, paulybrattka3cie}! \cite{goldwasser} discusses further cases.

Then, they are well-behaved under realizability: It is a common situation in computability and complexity theory that we have an algorithmic notion for some functions on some special sets $X$, $Y$ which we intend to lift to more general spaces $A$, $B$. We do this by fixing surjective encodings $\delta_A : \subseteq X \to A$, $\delta_B : \subseteq Y \to B$, and then calling e.g.~a function $f : A \to B$ computable, iff there is a computable function $F : \subseteq X \to Y$ such that the following diagram commutes:
$$\begin{CD}
X @>F>> Y\\
@VV\delta_AV @VV\delta_BV\\
A @>f>> B
\end{CD}$$
In general (depending on $\delta_A$, $\delta_B$), there will be algorithms (i.e.~functions $F : \subseteq X \to Y$) which do not compute any function $f : A \to B$, which leads to the canonization problem: The desire to find an algorithm $C_A : \subseteq X \to X$ with the properties $C_A(x) = C_A(y)$ whenever $\delta_A(x) = \delta_A(y)$, and $\delta_A(C_A(x)) = \delta_A(x)$. Note that in many cases canonization is known or suspected to be impossible with the available means.

On the other hand, every algorithm computes a multi-valued function, hence, the canonization problem is relegated to a far less fundamental position.

Algorithms lacking semantics as a function can nonetheless be very meaningful. A common example for this is the multi-valued function $\chi : \mathbb{R} \to \{0, 1\}$ with $0 \in \chi(x)$ iff $x \leq 1$ and $1 \in \chi(x)$ iff $x \geq 0$. $\chi$ is computable -- but the only computable functions from $\mathbb{R}$ to $\{0, 1\}$ are the constant ones. Hence, when working with real numbers, tests will have to be non-deterministic, i.e.~multi-valued functions. This then motivates an investigation of continuity for multivalued functions as in \cite{brattkahertling2,paulyziegler}.

Finally, as will be demonstrated in this paper, the properties of multi-valued functions have a nice impact on the degree-structure of many-one reductions: One always obtains a distributive lattice here.

An abridged version of the present article lacking proofs has appeared as \cite{paulysearchproblemscie}.

\section{Background}
Many-one reductions between multi-valued functions have been studied in complexity theory for several decades now, with complexity classes such as $\ppad$ \cite{papadimitrioub}, $\pls$ \cite{johnson} and $\fixp$ \cite{etessami} garnering a lot of attention. All three have a number of very interesting complete problems, often related to game theory. We just mention finding Nash equilibria in finite two player games with integer payoffs as a complete problem for $\ppad$ \cite{dengd}, finding Nash equilibria of generalized congestion games as complete for $\pls$ \cite{papadimitriouc} and finding exact Nash equilibria in finite three player games with integer payoffs as complete for $\fixp$ \cite{etessami}.

There also are a several problems which are known to be in both $\ppad$ and $\pls$, but where this is the best classification available. Computing winning strategies in parity or discounted payoff games is a typical example here (e.g.~\cite[Page 2]{paterson}). Despite this strong motivation to study $\ppad \cap \pls$, only in 2011 it was noticed (in a publication) that this class actually has complete problems \cite{papadimitrioue} - a fact that is an obvious consequence of the degree structure being a distributive lattice (which we show here). A systematic investigation of the degree structure seems to be missing so far.

In another setting for many-one reductions between multi-valued functions is the programme to classify the computational content of mathematical theorems in the Weihrauch lattice initiated in \cite{brattka3}, \cite{gherardi}. Here a mathematical theorem of the form \[ \forall x \in X \left (x \in D \Rightarrow \exists y \in Y \ T(x, y) \right )\]
is read as a multi-valued function $T : \subseteq X \mto Y$ with $\dom(T) = D$ which has to find a witness $y \in Y$ given some $x \in X$. The tool for classification is Weihrauch-reducibility, a form of many-one reducibility introduced originally in \cite{weihrauchb}, \cite{weihrauchc}.

Various theorems been classified in this framework: e.g.~the Hahn-Banach theorem \cite{gherardi}, Weak K\"onig's Lemma, the Intermediate Value theorem \cite{brattka2}, Nash's theorem on the existence of equilibria \cite{paulyincomputabilitynashequilibria}, Bolzano-Weierstrass \cite{gherardi4}, Brouwer's Fixed Point theorem \cite{paulybrattka3cie}, Ramsey's theorem (partially) \cite{shafer}, the existence of the Radon-Nikodym derivative \cite{hoyrup2b} and the Lebesgue Density Lemma \cite{hoelzl}.

Accompanying the investigation of specific degrees, also the overall degree structure has been studied. The Weihrauch degrees form a distributive lattice \cite{brattka2}, \cite{paulyreducibilitylattice}, and can be turned into a Kleene algebra when equipped with additional natural operations $\times$, $^*$ \cite{paulykojiro}. While some additional results in this area do depend on specific properties of Weihrauch reducibility, the fundamental ones only use generic properties of many-one reductions and multi-valued functions - and as such would also apply to the study of $\ppad$, $\pls$, etc.!

By outlining the properties of the category $\mult$ we simultaneously \emph{define} when a category is sufficiently similar to it to admit the same constructions as those used to study Weihrauch reducibility. In particular, we show how to introduce notions of many-one reductions in such a case, and derive the basic properties of the induced degree structures. The spirit of this endeavor bears similarity to the characterization of categories behaving like the one of partial functions as p-categories in \cite{robinson}, which also serves as our starting point.

As a means to study partial functions in an abstract setting, p-categories have largely been superseded by restriction categories introduced in \cite{cockett2,cockett3,cockett4}. The primary additional structure of a p-category is its (partial) product, whereas a restriction categories is built upon the mapping of a function to the identity restricted to the functions domain. Given the product, the restriction can be derived; and if a restriction category does admit a suitable product, then both settings are equivalent \cite[Theorem 5.2]{cockett4}. Despite restriction categories being the more general setting, we still use p-categories: Our definitions require the existence of the p-product anyway, whereas in the applications we have in mind the precise domain of a multivalued function is of less relevance.

Many-one reductions between sets are closely tied to pullbacks (e.g.~\cite[Subsection 6.6]{paulyreducibilitylattice}), this ceases to be true for multivalued functions. The reason for this lies in the fact that post-processing is included in our notion of reducibility for multivalued functions, whereas it generally is absent for sets.

While our goal of blending category and recursion theory has a significant history (e.g. \cite{longo}, \cite{dipaola}, see the survey \cite{cockett}), this does not include the study of reductions. On the other hand, categorical models of linear logic as studied in \cite{paiva},\cite{blass} admit certain similarities to the operations appearing in the present paper, however, have no strong connections to recursion or even complexity theory.

\section{Many-one reductions in a categorical setting}
\subsection{The category $\mult$}
It is easy to see that composition of multi-valued functions is associative, so they form a category $\mult$. One can lift disjoint unions and cartesian products from sets to multi-valued functions in the straight-forward way, we will denote the results by $f + g$ and $f \times g$. The disjoint union retains its r\^ole as the coproduct, however, the cartesian product is {\bf not} the categorical product. In the remainder of this subsection we shall explore some further specific properties of $\mult$, which, however, will not be needed for our development of many-one reduction. The required properties are then axiomatized in Subsections \ref{subsec:categories}, \ref{subsec:extensions}.

Like $\rel$, also $\mult$ does have a categorical product. Unlike in $\rel$, where the product is the same as the coproduct, in $\mult$ it is found in $A \otimes B := A + (A \times B) + B$ and projections $\pi_A : A \otimes B \to A$ defined via $\pi_A(a) = a$ for $a \in A$, $\pi_A((a,b)) = a$ for $(a,b) \in A \times B$ and $\pi_A(b) = \textsc{undef}$ for $b \in B$. Thus, $\mult$ essentially has the same products as the category $\parti$ of partial functions. For our purposes, $\otimes$ is badly behaved -- e.g.~it does not preserve computability.

A further -- decisive -- difference to the category $\rel$ of relations is the following:
\begin{proposition}
$\mult$ is not self-dual.
\begin{proof}[Sketch]
Assume that $\mult$ would admit a dual operation $^\dagger$. For reasons of cardinality, $^\dagger$ has to act like the identity on objects; in particular $^\dagger : \mult(2, 2) \to \mult(2,2)$. One can verify by exhausting all combinations that there is no self-dual operations on this hom-set with $(f \circ g)^\dagger = g^\dagger \circ f^\dagger$, thus $\mult$ cannot be self-dual.
\end{proof}
\end{proposition}

Both $\rel$ and $\mult$ can be obtained as the Kleisli-category of a monad acting on $\set$ -- and in both cases, the functor involved is the powerset functor $\mathcal{P}$. The difference lies in the multiplication: For $\rel$, it is $\mu_\rel : \mathcal{P}(\mathcal{P}(\mathbf{X})) \to \mathcal{P}(\mathbf{X})$ defined by $\mu_\rel(A) = \bigcup_{B \in A} B$. For $\mult$, it is $\mu_\mult : \mathcal{P}(\mathcal{P}(\mathbf{X})) \to \mathcal{P}(\mathbf{X})$ defined by $\mu_\mult(A) = \bigcup_{B \in A} B$ if $\emptyset \notin A$ and $\mu_\mult(A) = \emptyset$ if $\emptyset \in A$.(\footnote{The observation in this paragraph is due to an anonymous referee.})

\subsection{Categories of multivalued functions}
\label{subsec:categories}
As mentioned above, the starting point for our axiomatization are p-categories. \linebreak P-categories were introduced to capture the r\^ole of the cartesian product (i.e. the pairing function $\langle \ \rangle$) defined on partial functions, where it no longer coincides with the categorical product.

\begin{definition}[{\cite{robinson}}]
A p-category is a category $\mathcal{C}$ together with a naturally associative and naturally commutative bifunctor $\times : \mathcal{C} \times \mathcal{C} \to \mathcal{C}$ (the product\footnote{We hope that this nomenclature will not cause confusion, and point out again that the product in a p-category is not necessarily the product in the underlying category. It will be made clear whenever we refer to the categorical product instead.}), a natural transformation $\Delta$ (the diagonal) between the identity functor and the derived functor $X \mapsto X \times X$, and two families of natural transformations $(\pi_1^A)_{A \in Ob(\mathcal{C})}$ and $(\pi_2^B)_{B \in Ob(\mathcal{C})}$ (the projections) where $\pi_1^A$ is between the derived functor $X \mapsto X \times A$ and the identity, while $\pi_2^B$ is between the derived functor $X \mapsto B \times X$ and the identity, such that the following properties are given:
$$\begin{array}{ll}
\pi_1^X(X) \circ \Delta(X) = \pi_2^X(X) \circ \Delta(X) = \id_X & (\pi_1^Y(X) \times \pi_2^X(Y)) \circ \Delta(X \times Y) = \id_{X \times Y} \\
\pi_1^Y(X) \circ (\id_X \times \pi_1^Z(Y)) = \pi_1^{(Y \times Z)}(X) & \pi_1^Z(X) \circ (\id_X \times \pi_2^Y(Z)) = \pi_1^{(Y \times Z)}(X) \\
\pi_2^X(Y) \circ (\pi_1^Y(X) \times \id_Z) = \pi_2^{(X \times Y)}(Z) & \pi_2^X(Z) \circ (\pi_2^X(Y) \times \id_Z) = \pi_2^{(X \times Y)}(Z)
\end{array}$$
\end{definition}

For easier reading, we shall write $\pi_1^{X,Y}$ instead of $\pi_1^Y(X)$, $\pi_2^{X,Y}$ for $\pi_2^X(Y)$ and finally $\Delta_X$ in place of $\Delta(X)$. If the superscripts are obvious from the context, they may be dropped.

The treatment of partial maps in a categorical framework causes the concept of the domain of a map to split into two separate ones. With $\Dom(f)$ we denote the object $A$, if $f : A \to B$ is a morphism. Following \cite{dipaola}, we write $\dom(f)$ for the morphism $\pi_1^{A, B} \circ (\id_A \times f) \circ \Delta_A$, where $\pi_1$ is the first projection of the product $X \times Y$. One can interpret $\dom(f) : A \to A$ as the partial identity on that part of $A$ where the partial map $f$ is actually defined.

We point out that the composition $f \circ \dom(g)$ for suitable morphisms $f$, $g$ can be read as the restriction of $f$ to the domain of $g$. The concept of domains and restrictions allows us to formulate how projections work together with diagonals, as we find for any morphisms $f : X \to Y$, $g : X \to Z$:
$$\pi_2^{Y, Z} \circ \left (f \times g \right ) \circ \Delta_{X} = g \circ \dom(f) \ \ \textnormal{and} \ \ \pi_1^{Y, Z} \circ \left (f \times g \right ) \circ \Delta_{X} = f \circ \dom(g) $$

In a slightly more general situation, we can omit the diagonal and find:
$$\pi_1^{\codom(f), \codom(g)} \circ (f \times g) = f \circ \pi_1^{\Dom(f), \Dom(g)} \circ (\id_{\Dom(f)} \times \dom(g))$$

On morphisms of the form $\dom f : X \to X$, a partial order $\subseteq$ may be defined via $(\dom f) \subseteq (\dom g)$, if $(\dom f) \circ (\dom g) = \dom f$. This partial order even is a meet-semilattice, with composition $\circ$ taking the r\^ole of the infimum representing the intersection. For $f : A \to B$ and $g : A \to C$, the morphism $g\circ \dom(f)$ is the restriction of $g$ to the domain of $f$, so we could introduce the extension ordering for partial maps via $f \subseteq g$, if $f = g \circ \dom(f)$. Hence, any p-category may be considered to be poset enriched in a canonic way. In the general case, morphisms that are not domains will lack an infimum though, hence the canonic poset enrichment does not constitute a meet-semilattice enrichment.

Multivalued functions however have another preorder available, namely the \emph{easier than}-order $\preceq$ mentioned in the introduction. This does admit infima for arbitrary pairs $f, g : A \mto B$ of multivalued functions. Hence, we work in a meet-semilattice enriched p-category. Before we give a formal definition, we point out that the presence of coproduct is crucial for the later to be developed properties of the degree structures.

\begin{definition}
\label{def:multivaluedcategory}
A poset enriched p-category is a p-category $(\mathcal{C}, \times)$ together with a family $(\preceq_{A,B})_{A,B \in Ob(\mathcal{C})}$, where $\preceq_{A,B}$ is a preorder on the homsets $\mathcal{C}(A, B)$ that satisfies the following properties (we drop the subscript for $\preceq$ from here onwards):
\begin{enumerate}
\item $f_1 \preceq f_2$ implies $\left (g \circ f_1 \circ h \right ) \preceq \left (g \circ f_2 \circ h \right )$ for all suitable morphisms $f_1, f_2, g, h$
\item $f_1 \preceq f_2$ implies $f_1 \times g \preceq f_2 \times g$ for all suitable morphisms $f_1, f_2, g$
\item $f_\nu \preceq g_\nu$ for all $\nu < \alpha$ implies $\left ( \coprod_{\nu < \alpha} f_\nu \right ) \preceq \left ( \coprod_{\nu < \alpha} g_\nu \right )$ as long as these coproducts exist
\item $\dom(f) \subseteq \dom(g) \subseteq \dom(h)$ together with $\dom(h) \preceq \dom(f)$ implies $\dom(h) \preceq \dom(g)$
\end{enumerate}
We speak of a meet-semilattice enriched p-category, if additionally all preorders $\preceq$ admit infima compatible with composition and products, i.e.
\begin{enumerate}
\setcounter{enumi}{4}
\item $g \circ \inf \{f_1, f_2\} \circ h = \inf\{\left (g \circ f_1 \circ h \right ), \left (g \circ f_2 \circ h\right )\}$ for all suitable morphisms $f_1, f_2, g, h$
\item $g \times \inf \{f_1, f_2\} = \inf \{\left (g \times f_1\right), \left ( g \times f_2 \right )\}$
\item $\inf \{\left ( \coprod_{\nu < \alpha} f_\nu \right ), \left ( \coprod_{\nu < \alpha} g_\nu \right ) \} = \coprod_{\nu < \alpha} \inf \{f_\nu, g_\nu\}$ as long as these coproducts exist
\end{enumerate}
\end{definition}

The conditions 1.-3. and 5.-7. are straight-forward postulations that the local posets (meet-semilattice) are compatible with composition, the cartesian product and coproducts. Condition 4. can be read as stating that if the new preorder $\preceq$ behaves like the dual of the derived preorder $\subseteq$ somewhere, then it has to do so uniformly in a certain way. This is a technical criterion needed for our proofs. In the case that $\preceq$ is the \emph{easier than} preorder the conditions are modelled upon, the situation of condition 4. can only happen if all three domains coincide anyway. We point out that compatibility of $\preceq$ with composition and products already implies compatibility with domains:

\begin{proposition}
\label{proppreceqdomains}
For any morphisms $f, g : X \to Y$ in a poset enriched p-category, $f \preceq g$ implies $\dom(f) \preceq \dom(g)$.
\begin{proof}
From $f \preceq g$ we can conclude $(\id_X \times f) \preceq (\id_X \times g)$, which in turn implies: $$\left ( \pi_1^{X, Y} \circ (\id \times f) \circ \Delta_X \right ) \preceq \left ( \pi_1^{X, Y} \circ (\id \times g) \circ \Delta_X \right )$$ Evaluation of both sides of this statement yields the claim.
\end{proof}
\end{proposition}

We require two more conditions pertaining to the coproducts. For this, we fix some notation: Coproduct injections are denoted by $\iota_\mu^{(A_\nu)_{\nu<\alpha}} : A_\mu \to \left ( \coprod_{\nu < \alpha} A_\nu \right )$. The co-diagonal is written as $\nabla_A^\alpha : \left ( \coprod_{\nu < \alpha} A \right ) \to A$, it satisfies $\nabla_A^\alpha \circ \iota_\mu^{(A)_{\nu < \alpha}} = \id_A$ for all $\mu < \alpha$.

\begin{definition}
\label{defdistrcoprod}
A (poset enriched, meet-semilattice enriched) p-category has distributive coproducts of size $\alpha$, if all coproducts of size $\alpha$ exists in the underlying category, and there is a natural isomorphism $a : A \times \left ( \coprod_{\nu < \alpha} B_\nu \right ) \to \coprod_{\nu < \alpha} \left ( A \times B_\nu \right )$.
\end{definition}

\begin{definition}
A (poset enriched, meet-semilattice enriched) p-category is totally connected, if the underlying category is totally connected, i.e.~ if for any two objects $A, B \in Ob(\mathcal{C})$ there is a morphism $c_{A, B} : A \to B$. We assume\footnote{While this assumption makes $c_{\cdot,\cdot}$ behaving badly w.r.t.~equivalence of categories, it does simplify the proofs to follow. As the connectedness-structure is not particularly relevant here in its own right, this is a price we are willing to pay.} $c_{A,A} = \id_A$.
\end{definition}

The assumption of total connectedness for multivalued functions is based on the nowhere defined multivalued function $0 : X \to Y$ existing for any two objects. This family of morphisms has further interesting properties, which however are not relevant here. Now we can state and prove a number of useful properties of coproducts in our setting:

\begin{proposition}[{cf~\cite[Page 4]{cockett4}}]
\label{propretract}
In a totally connected (poset enriched, meet-semilattice enriched) p-category, coproduct injections are retractable.
\begin{proof}
A retract $\kappa_{\mu}^{(A_\nu)_{\nu <\alpha}} : \left ( \coprod_{\nu < \alpha} A_\nu \right ) \to A_\mu$ for the injection $\iota_{\mu}^{(A_\nu)_{\nu <\alpha}} : A_\mu \to \left ( \coprod_{\nu < \alpha} A_\nu \right )$ can be obtained as $\kappa_{\mu}^{(A_\nu)_{\nu <\alpha}} = \nabla_{A_\mu}^\alpha \circ \left ( \coprod_{\nu<\alpha} c_{A_\nu, A_\mu} \right )$. It is easy to see that $\kappa_{\mu}^{(A_\nu)_{\nu <\alpha}} \circ \iota_{\mu}^{(A_\nu)_{\nu <\alpha}} = \id_{A_\mu}$.
\begin{center}
\begin{tikzpicture}[node distance=1cm, auto]

  \node (A) {$A$};
  \node (AB) [below=of A] {$A \coprod B$};
  \node (AA) [right=2cm of AB] {$A \coprod A$};
  \draw[->] (A) to node {$\iota$} (AB);
  \draw[->] (AB) to node [swap] {$\id_A + c_{B,A}$} (AA);
  \draw[->] (AA) to node [swap] {$\nabla$} (A);
\end{tikzpicture}
\end{center}
\end{proof}
\end{proposition}

\begin{proposition}[{cf~\cite[Lemma 2.1]{cockett4}}]
\label{propdomnabla}
For any $X \in Ob(\mathcal{C})$ in a p-category with $\alpha$-coproducts we find $\dom(\nabla_X^\alpha) = \id_X$.
\begin{proof}
Abbreviate $\pi_2 := \pi_2^{X, \left (\coprod_{\nu <\alpha} X\right)}$. We have to show $\pi_2 \circ \left (\nabla_X^\alpha \times \id_{\left (\coprod_{\nu <\alpha} X\right)} \right ) \circ \Delta_{\left (\coprod_{\nu <\alpha} X\right)} = \id_X$. Due to the uniqueness condition for coproducts, this is equivalent to  $\left [ \pi_2 \circ \left (\nabla_X^\alpha \times \id_{\left (\coprod_{\nu <\alpha} X\right)} \right ) \circ \Delta_{\left (\coprod_{\nu <\alpha} X\right)} \right ] \circ \iota_\mu^{(X)_{\nu<\alpha}} = \iota_\mu^{(X)_{\nu<\alpha}}$ for all $\mu < \alpha$. By naturality of the diagonal and $\times$ being a functor, we obtain the equivalence to: $\pi_2 \circ \left (\id_X \times \iota_\mu^{(X)_{\nu<\alpha}} \right ) \circ \Delta_X = \iota_\mu^{(X)_{\nu<\alpha}}$, which is true.
\begin{center}
\begin{tikzpicture}[node distance=1cm, auto]
  \node (XX) {$X \coprod X$};
  \node (XXXX) [below=of XX] {$(X \coprod X) \times (X \coprod X)$};
  \node (XXX) [right=2cm of XXXX] {$X \times (X \coprod X)$};
  \draw[->] (XX) to node {$\Delta$} (XXXX);
  \draw[->] (XXXX) to node [swap] {$\nabla \times \id$} (XXX);
  \draw[->] (XXX) to node [swap] {$\pi_2$} (XX);
\end{tikzpicture}
\end{center}
\end{proof}
\end{proposition}

\begin{proposition}
\label{propdomiota}
For any family $(X_\nu)_{\nu < \alpha}$, $X_\nu \in Ob(\mathcal{C})$ in a p-category with $\alpha$-coproducts we find $\dom(\iota_\mu^{(X_\nu)_{\nu < \alpha}}) = \id_{X_\mu}$ for any $\mu < \alpha$.
\begin{proof}
The definition of coproducts requires $\left ( \coprod_{\nu < \alpha} \id_{X_\nu} \right ) \circ \iota_\mu^{(X_\nu)_{\nu < \alpha}} = \id_{X_\mu}$. Composition with $\dom(\iota_\mu^{(X_\nu)_{\nu < \alpha}})$ from the right yields $\left ( \coprod_{\nu < \alpha} \id_{X_\nu} \right ) \circ \iota_\mu^{(X_\nu)_{\nu < \alpha}} = \dom(\iota_\mu^{(X_\nu)_{\nu < \alpha}})$, comparison of the two right sides provides the claim.
\end{proof}
\end{proposition}

\begin{proposition}
\label{propcoproductdiags}
For families $(f_\nu : X_\nu \to Y_\nu)_{\nu < \alpha}$, $(g_\nu : X_\nu \to Z)_{\nu < \alpha}$ of morphisms in a p-category with distributive $\alpha$-coproducts the following identity holds:
$$\nabla_{\left ( \coprod_{\nu < \alpha} (Y_\nu \times Z)\right )}^\alpha \circ a \circ \left [ \left (\coprod_{\nu < \alpha} f_\nu \right ) \times \left (\coprod_{\mu < \alpha} g_\mu \right ) \right ] \circ \Delta_{\left ( \coprod_{\nu < \alpha} X_\nu \right )} = \coprod_{\nu < \alpha} \left ((f_\nu \times g_\nu) \circ \Delta_{X_\nu} \right )$$
where $a : \left [ \left (\coprod_{\nu < \alpha} Y_\nu \right ) \times \left ( \coprod_{\mu < \alpha} Z\right ) \right ] \to \left [ \coprod_{\mu < \alpha} \left (\coprod_{\nu < \alpha} (Y_\nu \times Z) \right ) \right ]$ is the canonic distributivity isomorphism.
\begin{proof}
We abbreviate $\nabla := \nabla_{\left ( \coprod_{\nu < \alpha} (Y_\nu \times Z)\right )}^\alpha$. It is sufficient to show that both sides of the equation are identical when composed with an arbitrary coproduct injection $\iota_{\eta}^{(X_\nu)_{\nu<\alpha}}$ for $\eta < \alpha$ from the right, hence we want to prove:
$$\nabla \circ a \circ \left [ \left (\coprod_{\nu < \alpha} f_\nu \right ) \times \left (\coprod_{\mu < \alpha} g_\mu \right ) \right ] \circ \Delta_{\left ( \coprod_{\nu < \alpha} X_\nu \right )} \circ \iota_{\eta}^{(X_\nu)_{\nu<\alpha}} = \iota_{\eta}^{(Y_\nu \times Z)_{\nu < \alpha}} \circ \left ( (f_\eta \times g_\eta) \circ \Delta_{X_\eta} \right )$$
As $\Delta$ is a natural transformation and $\times$ a functor, the injection $\iota_{\eta}^{(X_\nu)_{\nu<\alpha}}$ can be moved inwards to yield:
$$\nabla \circ a \circ \left [ (\iota_\eta^{(Y_\nu)_{\nu < \alpha}} \circ f_\eta) \times (\iota_\eta^{(Z)_{\nu < \alpha}} \circ g_\eta )\right ] \circ \Delta_{X_\eta} = \iota_{\eta}^{(Y_\nu \times Z)_{\nu < \alpha}} \circ \left ( (f_\eta \times g_\eta) \circ \Delta_{X_\eta} \right )$$
In the next step, the isomorphism $a$ is taken into account:
$$\nabla \circ \iota_{\eta}^{\left ( \coprod_{\nu<\alpha} (Y_\nu \times Z ) \right )_{\mu < \alpha}} \circ \iota_{\eta}^{(Y_\nu \times Z)_{\nu < \alpha}} \circ \left [  f_\eta \times g_\eta \right ] \circ \Delta_{X_\eta} = \iota_{\eta}^{(Y_\nu \times Z)_{\nu < \alpha}} \circ \left ( (f_\eta \times g_\eta) \circ \Delta_{X_\eta} \right )$$
The resulting identity follows directly from the definition of the co-diagonal.
\end{proof}
\end{proposition}

Occasionally we will be interested in special objects -- or morphisms that behave sufficiently like objects, i.e. domains -- in our categories. An initial object in a p-category is just an initial object in the underlying category, i.e. an object $I \in Ob(\mathcal{C})$ such that $|\mathcal{C}(I, A)| = 1$ for all $A \in Ob(\mathcal{C})$. We can generalize this notion to domains, by calling a domain $\dom i$ initial, if for any $A \in Ob(\mathcal{C})$ there is exactly one morphism $g$ with $g \circ \dom i = g$ and $\codom(g) = A$. Clear $I$ is initial, iff $\id_I = \dom \id_I$ is initial as a domain.

Usually an object $E$ in a category $\mathcal{C}$ is called \emph{empty}, if the existence of some morphism $g : A \to E$ implies that $A$ is an initial object. Clearly, this definition is somewhat contradicted with our requirement for the relevant categories to be totally connected -- only categories equivalent to the trivial category containing a single object and no further morphisms fulfills the criteria.

Calling a morphism $g$ in a p-category total, if $\dom(g) = \id_{\Dom(g)}$, we see that the total morphisms in a p-category form a sub-p-category, on which the p-category product even coincides with the categorical product. Now we define an empty object of a p-category to be an initial object in the underlying category which is empty in the subcategory of total morphisms. The concept of emptiness is extended to domains by calling an initial domain $\dom e$ empty, if for any total morphism $g$ with $\dom(e) \circ g = g$ we find $\Dom(g)$ to be an initial object.

Likewise, a final object of a p-category shall be a final object of the subcategory of total maps (cf.~\cite[Proposition 4.1]{cockett4}). A domain $\dom f$ is called final, if for any object $A$ there is exactly one total morphism $g$ with $\dom(f) \circ g = g$ and $\Dom(g) = A$. Now we may conclude that $E \times A$ is isomorphic to $E$ for any empty object $E$ and any object $A$ in a p-category, while for a final object $F$ we find $F \times A$ to be isomorphic to $A$. These properties can be extended to domains.

\subsection{Category extensions}
\label{subsec:extensions}
Many-one reductions (in a category theoretical framework) involve not only one, but two categories. For example, much of the classical complexity theory characterizes the degrees of \emph{computable} sets under \emph{polynomial-time computable} reductions. Likewise, computability theory looks at \emph{arbitrary sets} and \emph{computable} reductions. By moving to multivalued functions, we can drop the artificial structural dichotomy between the objects to be classified and the reduction witnesses, and find the latter to form a subcategory $\mathcal{S}$ of the former category $\mathcal{P}$. However, being a subcategory does not suffice: $\mathcal{S}$ and $\mathcal{P}$ will have to share all their remaining structure, too.

This situation is familiar in computable analysis \cite{weihrauchd}, unrelated to reductions: Here both computable as well as continuous maps (between represented spaces) form the ubiquitous foundation of the field, sharing the same structure. The impact of this on the categories underlying the field was noted in particular in \cite{bauer,bauer5}, see also \cite{pauly-synthetic}. Generalizing from computable to continuous usually just requires relativizing the proofs w.r.t.~an arbitrary oracle. Simply stating that the two categories have (to a large extent) the same structural properties does not do full justice to their relationship.

\begin{definition}
A category extension is a pair $(\mathcal{P}, \mathcal{S})$ of a category $\mathcal{P}$ and a wide subcategory $\mathcal{S} \subseteq \mathcal{P}$.
\begin{enumerate}
\item A category extension has coproducts, if $\mathcal{P}$ has coproducts and $\mathcal{S}$ is closed under the coproduct of $\mathcal{P}$.
\item A category extension is turned into a p-category extension, if $\mathcal{P}$ is equipped with a functor $\times$ turning it into a p-category, $\mathcal{S}$ is closed under $\times$ and a p-category with the restriction of $\times$. Moreover, we demand that $\mathcal{S}$ contains all domains of $\mathcal{P}$.
\item A p-category extension has distributive coproducts, if it has coproducts and the canonic distributivity morphism (Definition \ref{defdistrcoprod}) is present in $\mathcal{S}$.
\item A p-category extension is meet-semilattice enriched, if $\mathcal{P}$ is meet-semilattice enriched. This implies that $\mathcal{S}$ is poset-enriched.
\item A category extension is totally connected, if $\mathcal{S}$ is totally connected.
\item A domain is initial (empty, final) in a p-category extension $(\mathcal{P}, \mathcal{S})$, if it has this property in both $\mathcal{P}$ and $\mathcal{S}$.
\end{enumerate}
\end{definition}

\subsection{A generic definition of many-one reductions}
\label{subsecgeneric}
\begin{definition}
 A $(\alpha-)$\emph{many-one category extension} (moce) shall be a meet-semilattice enriched totally connected p-category extension with distributive coproducts (up to size $\alpha$).
\end{definition}

With the framework now established, we can proceed to define many-one reductions. There are two definitions of many-one reductions commonly found in the literature on search or function problems, which differ in the question whether the post-processing of the oracle answer still has access to the input. Forgetting the input leads to a simpler definition, and may make proofs of non-reducibility easier, while retaining it yields the nicer degree structure and allows to formulate stronger and more meaningful separation statements. We shall speak of strong many-one reductions if the original input is forgotten, and of many-one reductions otherwise.

Throughout this subsection, we assume that some $\alpha$-moce $(\mathcal{P}, \mathcal{S}, \times, \preceq)$ is given, with $\alpha \geq 2$, and refrain from mentioning it explicitly where this is unnecessary.

\begin{definition}[Strong many-one reductions]
Let $f \leq_{sm} g$ hold for $f, g \in \mathcal{P}$, if there are $H, K \in \mathcal{S}$ with $f \preceq H \circ g \circ K$.
\end{definition}

\begin{proposition}
\label{propsmpreorder}
$(\mathcal{P}, \leq_{sm})$ is a preordered class.
\begin{proof}
For any $f \in \mathcal{P}$, we have $\id_{\Dom(f)}, \id_{\codom(f)} \in \mathcal{S}$. Trivially, $f = \id_{\codom(f)} \circ f \circ \id_{\Dom(f)}$ holds. As $\preceq$ is a preorder, this implies $f \leq_{sm} f$.

Now assume $f \leq_{sm} g$ and $g \leq_{sm} h$ witnessed by $H, K, F, G \in \mathcal{S}$. Due to the assumptions on $\preceq$, $g \preceq F \circ h \circ K$ implies $H \circ g \circ K \preceq (H \circ F) \circ h \circ (G \circ K)$. Transitivity of $\preceq$ yields $f \preceq (H \circ F) \circ h \circ (G \circ K)$, hence $f \leq_{sm} h$ follows.
\end{proof}
\end{proposition}

\begin{definition}[Many-one reductions]
\label{defmanyonereductions}
Let $f \leq_{m} g$ hold for $f, g \in \mathcal{P}$, if there are $H, K \in \mathcal{S}$ with $f \preceq H \circ (\id_{\Dom(f)} \times (g \circ K)) \circ \Delta_{\Dom(f)}$.
\end{definition}

\begin{proposition}
\label{smimpliesm}
$f \leq_{sm} g$ implies $f \leq_m g$.
\begin{proof}
As we require $Ob(\mathcal{P}) = Ob(\mathcal{S})$, and $\mathcal{S}$ is closed under products, in particular we also have $\pi_2^{\Dom(f), \codom(g)} \in \mathcal{S}$ for the respective projection. Now assume $f \preceq H \circ g \circ K$. Then also $f \preceq (H \circ \pi_2^{\Dom(f), \codom(g)}) \circ (\id_{\Dom(f)} \times (g \circ K)) \circ \Delta_{\Dom(f)}$ is true.
\end{proof}
\end{proposition}

\begin{proposition}
\label{propleqmpreordered}
$(\mathcal{P}, \leq_{m})$ is a preordered class.
\begin{proof}
Reflexivity of $\leq_m$ follows from Propositions \ref{propsmpreorder}, \ref{smimpliesm}. Now assume $f \leq_m g$ witnessed by $F, G \in \mathcal{S}$, and $g \leq_m h$ witnessed by $H, K \in \mathcal{S}$. We abbreviate $\id_{Df} := \id_{\Dom(f)}$, $\id_{Dg} := \id_{\Dom(g)}$. Due to the assumptions on $\preceq$, $g \preceq H \circ (\id_{Dg} \times (h \circ K)) \circ \Delta_{\Dom(g)}$ implies: $$F \circ (\id_{Df} \times (g \circ G)) \circ \Delta_{\Dom(f)} \preceq F \circ (\id_{Df} \times (H \circ (\id_{Dg} \times (h \circ K)) \circ \Delta_{\Dom(g)} \circ G)) \circ \Delta_{\Dom(f)}$$ Using the transitivity of $\preceq$ and the naturality of the diagonal, one obtains: $$f \preceq F \circ (\id_{Df} \times (H \circ (G \times (h \circ K \circ G)) \circ \Delta_{\Dom(f)})) \circ \Delta_{\Dom(f)}$$
Now we use the distributivity of products over composition (i.e. the fact that $\times$ is a functor): $$f \preceq F \circ (\id_{Df} \times H) \circ (\id_{Df} \times ((G \times (h \circ K \circ G)) \circ \Delta_{\Dom(f)})) \circ \Delta_{\Dom(f)}$$
Then the associativity of products is used, with $a \in \mathcal{S}$ denoting the canonic isomorphism $a : ((A \times B) \times C) \to (A \times (B \times C))$ of suitable type:
$$f \preceq F \circ (\id_{Df} \times H) \circ a \circ (((\id_{Df} \times G) \circ \Delta_{\Dom(f)}) \times (h \circ K \circ G))  \circ \Delta_{\Dom(f)}$$
In the next step, again the distributivity of products over composition is relevant:
$$f \preceq F \circ (\id_{Df} \times H) \circ a \circ (((\id_{Df} \times G) \circ \Delta_{\Dom(f)}) \times \id_{\codom(h)}) \circ (\id_{Df} \times (h \circ K \circ G))  \circ \Delta_{\Dom(f)}$$
Now we abbreviate $M = F \circ (\id_{Df} \times H) \circ a \circ (((\id_{Df} \times G) \circ \Delta_{\Dom(f)}) \times \id_{\codom(h)})$ and $N = K \circ G$ and observe $M, N \in \mathcal{S}$, hence the following proves the remaining part of the claim:
$$f \preceq M \circ (\id_{\Dom(f)} \times (h \circ N)) \circ \Delta_{\Dom(f)}$$
\end{proof}
\end{proposition}

As usual, our interest is focused on the partial orders induced by the preorders (in particular by $\leq_m$) on their equivalence classes (or degrees). For any moce $(\mathcal{P}, \mathcal{S}, \times, \preceq)$, the partially ordered class of equivalence classes for $\leq_m$ shall be denoted by $\mathfrak{D}(\mathcal{P}, \mathcal{S}, \times, \preceq)$. The main result in this subsection is the following:

\begin{theorem}
\label{maintheolattice}
$\mathfrak{D}(\mathcal{P}, \mathcal{S}, \times, \preceq)$ is a distributive lattice.
\end{theorem}

The proof of Theorem \ref{maintheolattice} will be spread over the following lemmata, which also give category-theoretic descriptions of the suprema and infima in $\mathfrak{D}(\mathcal{P}, \mathcal{S}, \times, \preceq)$.

\begin{lemma}
\label{lemmacoproductsuprema}
$\mathfrak{D}(\mathcal{P}, \mathcal{S}, \times, \preceq)$ has $\alpha$-suprema, and these are given by $\alpha$-coproducts, i.e.:
$$\sup_{\leq_m, \nu < \alpha} f_\nu = \coprod_{\nu < \alpha} f_\nu$$
\begin{proof}
\begin{enumerate}
\item $f_\lambda \leq_{sm} \coprod_{\nu < \alpha} f_\nu$ for all $\lambda < \alpha$

By assumption, $\iota_\lambda^{(\Dom(f_\nu))_{\nu<\alpha}}, \kappa_\lambda^{(\codom(f_\nu))_{\nu<\alpha}} \in \mathcal{S}$ for the respective coproduct injections and the retracts of coproduct injections obtained via Proposition \ref{propretract}. The claim then follows from the following equation: $$f_\lambda = \kappa_\lambda^{(\codom(f_\nu))_{\nu<\alpha}} \circ \left ( \coprod_{\nu < \alpha} f_\nu \right ) \circ \iota_\lambda^{(\Dom(f_\nu))_{\nu<\alpha}}$$

\item $f_\lambda \leq_{m} \coprod_{\nu < \alpha} f_\nu$ for all $\lambda < \alpha$

Follows from 1. via Proposition \ref{smimpliesm}.

\item $\coprod_{\nu < \alpha} g \leq_{m} g$

Let $a : \left ( \coprod_{\nu < \alpha} \Dom(g) \right ) \times \codom(g) \to \Dom(g) \times \left ( \coprod_{\nu < \alpha} \codom(g) \right )$ be the canonic isomorphism due to the distributive nature of $\alpha$-coproducts and products\footnote{To be more precise, in order to construct this isomorphism we need to invoke the distributivity law for $\alpha$-coproducts and products twice, as well as the commutativity law for products.}. We abbreviate $\Delta := \Delta_{\left ( \coprod_{\nu < \alpha} \Dom(g)\right )}$; and use $\nabla^\alpha_{\Dom(g)}, a, \pi_2^{\Dom(g), \left ( \coprod_{\nu < \alpha} \codom(g) \right )} \in \mathcal{S}$. Hence, the following equation proves the claim:
$$\coprod_{\nu < \alpha} g = (\pi_2^{\Dom(g), \left ( \coprod_{\nu < \alpha} \codom(g) \right )} \circ a) \circ \left [\id_{\left ( \coprod_{\nu < \alpha} \Dom(g)\right )} \times \left (g \circ \nabla^\alpha_{\Dom(g)} \right ) \right ] \circ \Delta$$

\begin{center}
\begin{tikzpicture}[node distance=1cm, auto]
  \node (AA) {$A \coprod A$};
  \node (AAAA) [below=of AA] {$(A \coprod A) \times (A \coprod A)$};
  \node (AAA) [below=of AAAA] {$(A \coprod A) \times A$};
  \node (AAB) [right=2cm of AAA] {$(A \coprod A) \times B$};
  \node (ABB) [right=2cm of AAB] {$A \times (B \coprod B)$};
  \node (BB) [above=of ABB] {$B \coprod B$};
  \draw[->] (AA) to node {$\Delta$} (AAAA);
  \draw[->] (AAAA) to node {$\id \times \nabla$} (AAA);
  \draw[->] (AAA) to node {$\id \times g$} (AAB);
  \draw[->] (AAB) to node {$a$} (ABB);
  \draw[->] (ABB) to node {$\pi_2$} (BB);
  \draw[->] (AA) to node {$g+g$} (BB);
\end{tikzpicture}
\end{center}

To derive this equation, first consider the effect of the isomorphism $a$:
$$\coprod_{\nu < \alpha} g = \pi_2^{\Dom(g), \left ( \coprod_{\nu < \alpha} \codom(g) \right )} \circ \left [ \nabla^\alpha_{\Dom(g)}  \times \left (\coprod_{\nu < \alpha} g \right ) \right ] \circ \Delta$$

The interaction of projections and diagonals makes this equivalent to:
$$\coprod_{\nu < \alpha} g = \left (\coprod_{\nu < \alpha} g \right ) \circ \dom(\nabla^\alpha_{\Dom(g)})$$
The latter equation follows directly from Proposition \ref{propdomnabla}.

\item $f_\nu \leq_{m} g$ for all $\nu$ implies $\left ( \coprod_{\nu < \alpha} f_\nu \right ) \leq_{m} \left ( \coprod_{\nu < \alpha} g \right )$.

Let $f_\nu \leq_{m} g$ be witnessed by $H_\nu$, $K_\nu$. Further let $$a : \left (\coprod_{\nu < \alpha} \Dom(f_\nu) \right ) \times \left (\coprod_{\mu < \alpha} \codom(g) \right ) \to \coprod_{\mu < \alpha} \left [ \coprod_{\nu < \alpha} \left ( \Dom(f_\nu) \times \codom(g) \right ) \right ]$$
be the canonic isomorphism obtained from the distributivity law. We use $\nabla^\alpha$ to abbreviate $\nabla^\alpha_{\tiny \left [ \coprod_{\nu < \alpha} \left ( \Dom(f_\nu) \times \codom(g) \right ) \right ]}$. Then $\left ( \left (\coprod_{\nu < \alpha} H_\nu \right ) \circ \nabla^\alpha \circ a \right )$ and $\left ( \coprod_{\nu < \alpha} K_\nu \right )$ witness the claim. For this, consider:
$$\left ( \coprod_{\nu < \alpha} H_\nu \right ) \circ \nabla^\alpha \circ a \circ \left [ \id_{\left ( \coprod_{\nu < \alpha} \Dom(f_\nu) \right )} \times \left ( \left ( \coprod_{\nu < \alpha} g \right ) \circ \left ( \coprod_{\nu < \alpha} K_\nu \right ) \right ) \right ] \circ \Delta_{\left ( \coprod_{\nu < \alpha} \Dom(f_\nu) \right )}$$
Composition always commutes with coproducts of the same type:
 $$\left ( \coprod_{\nu < \alpha} H_\nu \right ) \circ \nabla^\alpha \circ a \circ \left [ \left ( \coprod_{\nu < \alpha} \id_{\Dom(f_\nu)} \right ) \times \left ( \coprod_{\nu < \alpha} (  g  \circ  K_\nu ) \right ) \right ] \circ \Delta_{\left ( \coprod_{\nu < \alpha} \Dom(f_\nu) \right )}$$
 Now we can invoke Proposition \ref{propcoproductdiags} to obtain:
 $$ \left ( \coprod_{\nu < \alpha} H_\nu \right ) \circ \left [ \coprod_{\nu < \alpha} \left ( (\id_{\Dom(f_\nu)} \times (g \circ K_\nu)) \circ \Delta_{\Dom(f_\nu)} \right ) \right ] $$

Invoking commutativity of coproducts and composition again, we get:
$$\coprod_{\nu < \alpha} \left [ H_\nu \circ \left (\id_{\Dom(f_\nu)} \times (g \circ K_\nu) \right) \circ \Delta_{\Dom(f)}\right ]$$
As $\preceq$ and $\alpha$-coproducts commute, we know that $\left ( \coprod_{\nu < \alpha} f_\nu \right )$ is $\preceq$-below the expression above. This concludes this part of the proof.

\item $f_\nu \leq_{m} g$ for all $\nu$ implies $\left ( \coprod_{\nu < \alpha} f_\nu \right ) \leq_{m} g$.

This follows by applying transitivity of $\leq_m$ from Proposition \ref{propleqmpreordered} to 3. and 4.

\item The claim is equivalent to 2. and 5.
\end{enumerate}
\end{proof}
\end{lemma}

The infima are not given by a purely category-theoretic construction, but rather rely on the poset enrichment together with the assumption that in $\mathcal{P}$, suitable binary infima actually exist. As projections and injections are all included in $\mathcal{S}$ by assumption, we can define the following:
\begin{definition}
\label{defoplus}
For any morphisms $f, g \in \mathcal{P}$ define $(f \oplus g) : (\Dom(f) \times \Dom(g)) \to (\codom(f) \coprod \codom(g))$ via: $$(f \oplus g) = \inf_{\preceq, i \in \{1, 2\}} \{\iota_i^{\codom(f), \codom(g)} \circ \pi_i^{\codom(f), \codom(g)} \} \circ (f \times g)$$
\end{definition}

\begin{lemma}
$\mathfrak{D}(\mathcal{P}, \mathcal{S}, \times, \preceq)$ has (binary) infima, and these are given by $\oplus$, i.e.: $$\inf_{\leq_m} \{f, g\} = f \oplus g$$
\begin{proof}
\begin{enumerate}
\item $(f \oplus g) \leq_{sm} f$ and $(f \oplus g) \leq_{sm} g$

As $\oplus$ inherits commutativity from products and coproducts, it is sufficient to prove $(f \oplus g) \leq_{sm} f$. This is witnessed by the morphisms $\iota_1^{\codom(f), \codom(g)}$ and \linebreak $\left [ \pi_1^{\Dom(f), \Dom(g)} \circ (\id_{\Dom(f)} \times \dom(g)) \right ]$, as we find: $$\begin{array}{cl} & \iota_1^{\codom(f), \codom(g)} \circ f \circ \left [ \pi_1^{\Dom(f), \Dom(g)} \circ (\id_{\Dom(f)} \times \dom(g)) \right ] \\ = & \iota_1^{\codom(f), \codom(g)} \circ \pi_1^{\codom(f), \codom(g)} \circ (f \times g)\end{array}$$ The latter expression is clearly $\preceq$-above $f \oplus g$, as can be verified from Definition \ref{defoplus}.
\begin{center}
\begin{tikzpicture}[node distance=1cm, auto]
  \node (XA) {$X \times A$};
  \node (XA2) [below=of XA]{$X \times A$};
  \node (YB) [right=2cm of XA] {$Y \times B$};
  \node (X) [right=2cm of XA2] {$X$};
  \node (Y) [right=2cm of X] {$Y$};
  \node (YBp) [right=of Y] {$Y$};
  \draw[->,dashed] (XA) to node {$\id \times \dom(g)$} (XA2);
  \draw[->] (XA) to node {$f \times g$} (YB);
  \draw[->] (XA2) to node {$\pi$} (X);
  \draw[->] (X) to node {$f$} (Y);
  \draw[->] (Y) to node {$\iota$} (YBp);
  \draw[->] (YB) to node {$\pi$} (Y);
\end{tikzpicture}
\end{center}
\item $(f \oplus g) \leq_{m} f$, $(f \oplus g) \leq_{m} g$

Follows from 1. via Proposition \ref{smimpliesm}.

\item If $h \leq_m f$ and $h \leq_m g$, then $h \leq_m (f \oplus g)$.

Let $h \leq_m f$ be witnessed by $H_1$, $K_1$ and let $h \leq_m g$ be witnessed by $H_2$, $K_2$. Further let {\small $a : \left [ \Dom(h) \times \left (\codom(f) \coprod \codom(g)\right ) \right ] \to \left [ \left ( \Dom(h) \times \codom(f) \right ) \coprod \left ( \Dom(h) \times \codom(g) \right ) \right ]$} be the canonic distributivity isomorphism. We abbreviate $\iota_i := \iota_i^{\codom(f), \codom(g)}$, $\pi_i := \pi_i^{\codom(f), \codom(g)}$.

The claim now is witnessed by $\left [ \nabla_{\codom(h)} \circ \left (\coprod_{i \in \{1, 2\}} H_i \right ) \circ a \right ]$ and $\left [ \left (K_1 \times K_2 \right ) \circ \Delta_{\Dom(h)}\right ]$. To prove this, we have to show that the following morphism is $\preceq$-above $h$:
{\tiny $$\left [ \nabla_{\codom(h)} \circ \left (\coprod_{i \in \{1, 2\}} H_i \right ) \circ a \right ] \circ \left [ \id_{\Dom(h)} \times \left ( \inf_{\preceq, i \in \{1, 2\}} \{\iota_i \circ \pi_i \} \circ (f \times g) \circ \left [ \left (K_1 \times K_2 \right ) \circ \Delta_{\Dom(h)}\right ] \right ) \right ] \circ \Delta_{\Dom(h)}$$ }
As $\inf$ and $\circ$ are compatible, we can move the $\inf$-operator to the outside, and obtain a morphism that is $\preceq$-below the preceding one, hence it will suffice to show that $h$ is $\preceq$-below the following:
{ \tiny $$\inf_{\preceq, i \in \{1, 2\}} \left \{ \nabla_{\codom(h)} \circ \left (\coprod_{j \in \{1, 2\}} H_j \right ) \circ a \circ \left [ \id_{\Dom(h)} \times \left ( \iota_i \circ \pi_i \circ \left [ (f \circ K_1) \times (g \circ K_2) \right ] \circ \Delta_{\Dom(h)} \right ) \right ] \circ \Delta_{\Dom(h)} \right \}$$ }
Using the standard properties of the isomorphism $a$, coproducts, injections and the co-diagonal, this is equivalent to:
$$\inf_{\preceq, i \in \{1, 2\}} \left \{ H_i \circ \left [ \id_{\Dom(h)} \times \left ( \pi_i \circ \left [ (f \circ K_1) \times (g \circ K_2) \right ] \circ \Delta_{\Dom(h)}\right ) \right ] \circ \Delta_{\Dom(h)} \right \}$$
Applying the projections yields the following equivalent expression:
{ \small $$\inf_{\preceq} \left \{ \begin{array}{l} \left [ H_1 \circ  \left ( \id_{\Dom(h)} \times \left [ f \circ K_1 \circ \dom(g \circ K_2)\right ] \right ) \circ \Delta_{\Dom(h)} \right ], \\ \left [ H_2 \circ \left ( \id_{\Dom(h)} \times \left [ g \circ K_2 \circ \dom(f \circ K_1)\right ] \right ) \circ \Delta_{\Dom(h)} \right ] \end{array}\right \}$$ }
The domain-morphisms can be moved past the diagonal to arrive at:
{ \small $$\inf_{\preceq} \left \{ \begin{array}{l} \left [ H_1 \circ  \left ( \id_{\Dom(h)} \times \left [ f \circ K_1\right ] \right ) \circ \Delta_{\Dom(h)} \right ] \circ \dom(g \circ K_2), \\ \left [ H_2 \circ \left ( \id_{\Dom(h)} \times \left [ g \circ K_2 \right ] \right ) \circ \Delta_{\Dom(h)} \right ] \circ \dom(f \circ K_1) \end{array} \right \}$$ }
By assumption, we have $h \preceq \left [ H_2 \circ \left ( \id_{\Dom(h)} \times \left [ g \circ K_2 \right ] \right ) \circ \Delta_{\Dom(h)} \right ]$, so from Proposition \ref{proppreceqdomains} we can conclude $\dom(h) \preceq \dom(\left [ H_2 \circ \left ( \id_{\Dom(h)} \times \left [ g \circ K_2 \right ] \right ) \circ \Delta_{\Dom(h)} \right ])$. Straight-forward consideration shows $\dom(\left [ H_2 \circ \left ( \id_{\Dom(h)} \times \left [ g \circ K_2 \right ] \right ) \circ \Delta_{\Dom(h)} \right ]) \subseteq \dom(g \circ K_2)$. By composition with $\dom(h)$ from the right, we arrive at $$\dom(\left [ H_2 \circ \left ( \id_{\Dom(h)} \times \left [ g \circ K_2 \right ] \right ) \circ \Delta_{\Dom(h)} \right ] \circ \dom(h)) \subseteq \dom(g \circ K_2 \circ \dom(h)) \subseteq \dom(h)$$ and $\dom(h) \preceq \dom(\left [ H_2 \circ \left ( \id_{\Dom(h)} \times \left [ g \circ K_2 \right ] \right ) \circ \Delta_{\Dom(h)} \right ] \circ \dom(h))$, so \ref{def:multivaluedcategory} (4) implies $\dom(h) \preceq \dom(g \circ K_2 \circ \dom(h))$. If $H_2$, $K_2$ witnesses $h \preceq g$, then so do $H_2$, $(K_2 \circ \dom(h))$, so w.l.o.g. we may assume $\dom(g \circ K_2 \circ \dom(h)) = \dom(g \circ K_2)$, so we even find $\dom(h) \preceq \dom(g \circ K_2)$. But then $h \preceq \left [ H_1 \circ  \left ( \id_{\Dom(h)} \times \left [ f \circ K_1\right ] \right ) \circ \Delta_{\Dom(h)} \right ]$ implies: $$h \preceq \left [ H_1 \circ  \left ( \id_{\Dom(h)} \times \left [ f \circ K_1\right ] \right ) \circ \Delta_{\Dom(h)} \right ] \circ \dom(g \circ K_2)$$ By symmetry, we can use the same way to prove: $$h \preceq \left [ H_2 \circ \left ( \id_{\Dom(h)} \times \left [ g \circ K_2 \right ] \right ) \circ \Delta_{\Dom(h)} \right ] \circ \dom(f \circ K_1)$$
This concludes the proof of the claim.

\item 2. and 3. are the defining properties of the infimum.
\end{enumerate}
\end{proof}
\end{lemma}

\begin{lemma}
$\mathfrak{D}(\mathcal{P}, \mathcal{S}, \times, \preceq)$ is distributive, i.e. for all $f, g_\nu \in \mathcal{P}$:

$$f \oplus \left (\coprod_{\nu < \alpha} g_\nu \right ) \equiv_{m} \coprod_{\nu < \alpha} (f \oplus g_\nu)$$
\begin{proof}
\begin{enumerate}
\item $f \oplus \left (\coprod_{\nu < \alpha} g_\nu \right ) \leq_{sm} \coprod_{\nu < \alpha} (f \oplus g_\nu)$

Let $a : \left [ \Dom(f) \times \left (\coprod_{\nu < \alpha} \Dom(g_\nu)\right ) \right ] \to \left [ \coprod_{\nu < \alpha} (\Dom(f) \times \Dom(g_\nu)) \right ]$ be the canonic distributivity isomorphism. Further consider the canonic associativity isomorphism $b : \left [ \coprod_{\nu < \alpha} \left ( \codom(f) \coprod \codom(g_\nu) \right ) \right ] \to \left [ \codom(f) \coprod \left ( \coprod_{\nu < \alpha} \codom(g_\nu)\right ) \right ]$. Then $a$ and $b$ witness the reduction, i.e.: $$\left [ f \oplus \left (\coprod_{\nu < \alpha} g_\nu \right )\right ] \preceq b \circ \left [ \coprod_{\nu < \alpha} (f \oplus g_\nu)\right ]  \circ a$$
We abbreviate $\pi_i :=\pi_i^{\codom(f), \codom(g_\mu)}$ and $\iota_i := \iota_i^{\codom(f), \left (\coprod_{\nu < \alpha} \codom(g_\nu) \right )}$.
As $a$ is an isomorphism, we can apply the inverse $a^{-1}$ from the right on both sides and obtain an equivalent statement:
$$\left [ f \oplus \left (\coprod_{\nu < \alpha} g_\nu \right )\right ] \circ a^{-1} \preceq b \circ \left [ \coprod_{\nu < \alpha} (f \oplus g_\nu)\right ] $$
Now both sides have a coproduct as a domain, so by compatibility of $\preceq$ and coproducts as well as composition, the statement above is equivalent to the one below holding for all $\mu < \alpha$:
$$\left [ f \oplus \left (\coprod_{\nu < \alpha} g_\nu \right )\right ] \circ a^{-1} \circ \iota_\mu^{(\Dom(f) \times \Dom(g_\nu))_{\nu < \alpha}} \preceq b \circ \left [ \coprod_{\nu < \alpha} (f \oplus g_\nu)\right ] \circ \iota_\mu^{(\Dom(f) \times \Dom(g_\nu))_{\nu < \alpha}} $$
This evaluates to:
$$\left [ f \oplus \left (\coprod_{\nu < \alpha} g_\nu \right )\right ] \circ (\id_{\Dom(f)} \times \iota_\mu^{(\Dom(g_\nu))_{\nu < \alpha}}) \preceq b \circ \iota_{\mu}^{(\codom(f) \coprod \codom(g_\nu))_{\nu < \alpha}} \circ (f \oplus g_\mu)  $$

To prove this statement, we insert the definition of $\oplus$ and move the coproduct injection on the left side further to the left:
{\small $$\begin{array}{c} \inf \limits_{\preceq, i \in \{1, 2\}} \{\iota_i \circ \pi_i^{\codom(f), \left (\coprod_{\nu < \alpha} \codom(g_\nu) \right )} \} \circ (\id_{\codom(f)} \times \iota_\mu^{(\codom(g_\nu))_{\nu < \alpha}}) \circ \left [ f \times  g_\mu \right ] \\ \preceq \\ b \circ \iota_{\mu}^{(\codom(f) \coprod \codom(g_\nu))_{\nu < \alpha}} \circ  \inf_{\preceq, i \in \{1, 2\}} \{\iota_i^{\codom(f), \codom(g_\mu)} \circ \pi_i \}  \circ  (f \times g_\mu)\end{array}$$ }
As we assume that composition preserves infima, we can move these to the outside. Additionally, we can drop the composition with $(f \times g_\mu)$ from both sides to arrive at a stronger statement:
{\small $$\begin{array}{c} \inf \limits_{\preceq, i \in \{1, 2\}} \left \{\iota_i \circ \pi_i^{\codom(f), \left (\coprod_{\nu < \alpha} \codom(g_\nu) \right )} \circ (\id_{\codom(f)} \times \iota_\mu^{(\codom(g_\nu))_{\nu < \alpha}}) \right \}  \\ \preceq \\  \inf \limits_{\preceq, i \in \{1, 2\}} \left \{ b \circ \iota_{\mu}^{(\codom(f) \coprod \codom(g_\nu))_{\nu < \alpha}} \circ \iota_i^{\codom(f), \codom(g_\mu)} \circ \pi_i \right \}\end{array}$$ }
On the left side, we can move the projections past the product due to $\dom(\iota_\mu^{(\codom(g_\nu))_{\nu < \alpha}}) = \id_{\codom(g_\mu)}$ (Proposition \ref{propdomiota}), and on the right side we take into consideration the effects of the isomorphism $b$:
$$\inf \limits_{\preceq} \left \{ \left [\iota_1 \circ \pi_1 \right ], \left [ \iota_2 \circ \iota_\mu^{(\codom(g_\nu))_{\nu < \alpha}} \circ \pi_2 \right ] \right \}  \preceq  \inf \limits_{\preceq} \left \{ \left [\iota_1 \circ \pi_1 \right ], \left [ \iota_2 \circ \iota_\mu^{(\codom(g_\nu))_{\nu < \alpha}} \circ \pi_2 \right ] \right \}$$
As both sides are identical, this statement true, and, as shown above, implies our claim.

\item $f \oplus \left (\coprod_{\nu < \alpha} g_\nu \right ) \leq_{m} \coprod_{\nu < \alpha} (f \oplus g_\nu)$

Follows from 1. via Proposition \ref{smimpliesm}.

\item $\coprod_{\nu < \alpha} (f \oplus g_\nu) \leq_{m} f \oplus \left (\coprod_{\nu < \alpha} g_\nu \right )$

This direction holds in every lattice, hence, it follows from the first part of Theorem \ref{maintheolattice}.
\end{enumerate}
\end{proof}
\end{lemma}

The presence of special objects in a moce $(\mathcal{P}, \mathcal{S}, \times, \preceq)$ implies the existence of special degrees in $\mathfrak{D}(\mathcal{P}, \mathcal{S}, \times, \preceq)$, as will be elaborated next:

\begin{proposition}
\label{propinitialbot}
If $(\mathcal{P}, \mathcal{S}, \times, \preceq)$ has an initial domain, then $\mathfrak{D}(\mathcal{P}, \mathcal{S}, \times, \preceq)$ has a bottom element.
\begin{proof}
Let $i = \dom i : I \to I$ be an initial domain in $\mathcal{S}$. We claim $i \leq_{sm} f$ for any $f \in \mathcal{P}$. By Proposition \ref{smimpliesm} this implies the original statement. As $\mathcal{S}$ is totally connected, there is some morphism $c_{\codom(f), I} : \codom(f) \to I$ in $\mathcal{S}$. Also, there is a morphism $c_{I, \Dom(f)} : I \to \Dom(f)$ satisfying $c_{I, \Dom(f)} \circ i = c_{I, \Dom(f)}$, as $i$ is initial. Then $c_{\codom(f), I} \circ f \circ c_{I, \Dom(f)} : I \to I$ must be equal to $i$, as we have both $i \circ i = i$ as well as $c_{\codom(f), I} \circ f \circ c_{I, \Dom(f)} \circ i = c_{\codom(f), I} \circ f \circ c_{I, \Dom(f)}$.
\end{proof}
\end{proposition}

\subsection{The Kleene-algebra of many-one degrees}
\label{subseckleene}
Besides the lattice-structure, the many-one degrees also have the structure of a Kleene-algebra\footnote{We are referring to the algebraic concept here, not to the distributive lattice with an involution!}\cite{kozen}, provided certain conditions are fulfilled. We shall start by discussion the underlying idempotent semiring. The addition in the Kleene-algebra is the supremum of the lattice, i.e. the coproduct. The multiplication in the Kleene-algebra is induced by the product of the p-category, as to be shown next.

\begin{lemma}
\label{lemmaproductdegrees}
$\times$ induces an operation on $\mathfrak{D}(\mathcal{P}, \mathcal{S}, \times, \preceq)$, i.e. $f_i \leq_m g_i$ for $i \in \{1, 2\}$ implies $(f_1 \times f_2) \leq_m (g_1 \times g_2)$.
\begin{proof}
Let $f_i \leq_m g_i$ be witnessed by $H_i$, $K_i$, and let \[\begin{array}{rcl} a & : & \left [ \left ( \Dom(f_1) \times \Dom(f_2) \right ) \times  \left (\codom(g_1) \times \codom(g_2) \right ) \right ] \\ & \to & \left [ \left (\Dom(f_1) \times \codom(g_1) \right ) \times \left (\Dom(f_2) \times \codom(g_2) \right ) \right ]\end{array}\] be the canonic isomorphism constructed from associativity and commutativity of $\times$. Then $(H_1 \times H_2) \circ a$ and $(K_1 \times K_2)$ witnesses $(f_1 \times f_2) \leq_m (g_1 \times g_2)$ as can be seen from $$(H_1 \times H_2) \circ a \circ \left [ \id_{\Dom(f_1) \times \Dom(f_2)} \times \left ((g_1 \times g_2) \circ (K_1 \times K_2) \right ) \right ] \circ \Delta_{\Dom(f_1) \times \Dom(f_2)}$$
being equal to
$$\left [H_1 \circ \left (\id_{\Dom(f_1)} \times (g_1 \circ K_1)\right) \circ \Delta_{\Dom(f_1)}\right ] \times \left [H_2 \circ \left (\id_{\Dom(f_2)} \times (g_2 \circ K_2)\right) \circ \Delta_{\Dom(f_2)}\right ]$$ together with the fact that $\times$ is compatible with $\preceq$.
\end{proof}
\end{lemma}

For completeness, we shall also state the following, while omitting the trivial proof:
\begin{lemma}
$\times$ induces an operation on the $\leq_{sm}$-degrees, i.e. $f_i \leq_{sm} g_i$ for $i \in \{1, 2\}$ implies $(f_1 \times f_2) \leq_{sm} (g_1 \times g_2)$.
\end{lemma}

\begin{theorem}
Let the p-category $\mathcal{P}$ have an empty domain $e$ and a final domain $f$. Then the degrees $\mathfrak{D}(\mathcal{P}, \mathcal{S}, \times, \preceq)$ together with the operations $\coprod$ and $\times$ and the induced constants $\mathbf{e}, \mathbf{f} \in \mathfrak{D}$ form an idempotent commutative semiring, i.e. the following hold for all $\mathbf{a}, \mathbf{b}, \mathbf{c} \in \mathfrak{D}$:
\begin{enumerate}
\item $\mathbf{a} \coprod \mathbf{a} = \mathbf{a}$, $(\mathbf{a} \coprod \mathbf{b}) \coprod \mathbf{c} = \mathbf{a} \coprod (\mathbf{b} \coprod \mathbf{c})$, $\mathbf{a} \coprod \mathbf{b} = \mathbf{b} \coprod \mathbf{a}$
\item $\mathbf{a} \coprod \mathbf{e} = \mathbf{a}$
\item $(\mathbf{a} \times \mathbf{b}) \times \mathbf{c} = \mathbf{a} \times (\mathbf{b} \times \mathbf{c})$, $\mathbf{a} \times \mathbf{b} = \mathbf{b} \times \mathbf{a}$
\item $\mathbf{a} \times \mathbf{f} = \mathbf{a}$, $\mathbf{a} \times \mathbf{e} = \mathbf{e}$
\item $\mathbf{a} \times (\mathbf{b} \coprod \mathbf{c}) = (\mathbf{a} \times \mathbf{b}) \coprod (\mathbf{a} \times \mathbf{c})$
\end{enumerate}
\begin{proof}
First of all, note that by Lemmata \ref{lemmacoproductsuprema}, \ref{lemmaproductdegrees} the operations are well-defined. Further, by Proposition \ref{smimpliesm} morphisms in $\mathcal{P}$ that are isomorphic over $\mathcal{S}$ represent the same degree in $\mathfrak{D}(\mathcal{P}, \mathcal{S}, \times, \preceq)$. Hence, 1. follows from Lemma \ref{lemmacoproductsuprema}. As every empty domain is initial,  2. follows from Lemma \ref{lemmacoproductsuprema} and Proposition \ref{propinitialbot}. Claim 3. is a direct consequence of associativity and commutativity of the functor $\times$, while 4. comes from the interaction of $\times$ with special objects or domains. Finally, 5. is implied by the requirement that the functor $\times$ distributes over coproducts.
\end{proof}
\end{theorem}

The product can be iterated: We inductively define $f^1 = f$, $f^{n+1} = f^n \times f$ for any morphism, and follow by $f^* = \coprod_{n \in \mathbb{N}} f^n$, assuming this coproduct exists. Intuitively, access to $f^*$ allows us to use $f$ for any predetermined finite number of times in parallel; hence, the operation $^*$ is suitable to introduce a generalization of $wtt$-degrees in our framework.

In a $\aleph_0$-moce, we have the prerequisites to deal with all countable coproducts, in particular with $f^*$. However, our computability-inspired examples do not fulfill the respective criteria: The category of computable functions is not closed under countable coproducts. However, forming countable coproducts is usually unproblematic in the super-category $\mathcal{P}$, and sufficiently uniform countable coproducts, such as $f^*$ even preserve computability. (Alternatively, a formulation using internal category theory inside the effective topos should resolve the issue.) We will prove the existence of the Kleene-algebra structure assuming that $\mathcal{S}$ has all needed coproducts; so in order to apply the result, their presence has to be checked individually.

As preparation, we point out that for any two objects $X, Y \in Ob(\mathcal{S})$ we find $X^* \times Y^*$ to be isomorphic to $(X \times Y)^* \coprod \left ( \coprod_{n, m \in \mathbb{N}, n \neq m} (X^n \times Y^m) \right )$, hence, we have a retractable embedding $(X \times Y)^* \hookrightarrow (X^* \times Y^*)$ in $\mathcal{S}$ due to the assumption that $\mathcal{S}$ is totally connected. Further, we need to consider objects of the form $(X^*)^m$. Iterating the distributivity of products over coproducts, we obtain an isomorphism $a_m : (X^*)^m \to \coprod_{n_1, \ldots, n_m \in \mathbb{N}} X^{n_1 + \ldots + n_m}$ in $\mathcal{S}$. For any $n \in \mathbb{N}$, we let $p(n)$ denote the number of distinct order-depending summations of the form $n = n_1 + \ldots n_m$ with varying $m \in \mathbb{N}$. Then we obtain an isomorphism $a : (X^*)^* \to \coprod_{n \in \mathbb{N}} \left (\coprod_{i \leq p(n)} X^n\right )$ in $\mathcal{S}$ by taking first the coproduct over all $a_m$, and then rearranging the resulting coproduct. Motivated by this, we will use $\nabla_X^* : (X^*)^* \to X^*$ to denote the composition $\nabla_X^* = \left ( \coprod_{n \in \mathbb{N}} \nabla_X^{p(n)} \right ) \circ a$. Furthermore, consider the product $\left (\coprod_{n \in \mathbb{N}} \left (\coprod_{i \leq p(n)} X^n\right ) \right ) \times \left (\coprod_{m \in \mathbb{N}} Y^m\right )$. Due to distributivity and associativity, this is isomorphic to: \[\left [\coprod_{n \in \mathbb{N}} \coprod_{i \leq p(n)} (X^n \times Y^n) \right ] \coprod \left [ \coprod_{n, m \in \mathbb{N}, n\neq m} \coprod_{i \leq p(n)} (X^n \times Y^m)\right ]\] Compose the respective isomorphisms with a retract to the first component of the final coproduct to obtain a canonic morphism $c : (X^*)^* \times Y^* \to \left [\coprod_{n \in \mathbb{N}} \coprod_{i \leq p(n)} (X^n \times Y^n) \right ]$. Finally, apply $\left ( \coprod_{n \in \mathbb{N}} \pi_1^{X^n, Y^n} \right )$ and another isomorphism to obtain the canonic morphism $\chi_{X, Y} : ((X^*)^* \times Y^*) \to (Y^*)^*$.

\begin{theorem}
\label{theoclosureoperator}
Given an $\aleph_0$-moce $(\mathcal{P}, \mathcal{S}, \times, \preceq)$, the operation $^*$ induces a closure operator on $\mathfrak{D}(\mathcal{P}, \mathcal{S}, \times, \preceq)$, i.e. for all $f, g \in \mathcal{P}$:\begin{enumerate}
\item $f \leq_m f^*$
\item $f \leq_m g$ implies $f^* \leq_m g^*$
\item $(f^*)^* \leq_m f^*$
\end{enumerate}
\begin{proof}
\begin{enumerate}
\item $f \leq_m f^*$

This follows directly from Lemma \ref{lemmacoproductsuprema} (2) together with the definition of $^*$.

\item $f \leq_m g$ implies $f^* \leq_m g^*$

Let $f \leq_m g$ be witnessed by $H, K \in \mathcal{S}$. Let $a \in \mathcal{S}$ be an retract of the embedding $(\Dom(f)^* \times \codom(g)^*) \hookrightarrow (\Dom(f) \times \codom(g))^*$. Then $(H^* \circ a)$ and $K^*$ witness the claim.

\item $(f^*)^* \leq_m f^*$

The witnesses are $\nabla_{\Dom(f)}^*$ and $\chi_{\Dom(f), \codom(f)}$.
\end{enumerate}
\end{proof}
\end{theorem}

\begin{corollary}
Let the p-category $\mathcal{P}$ have an empty object $E$ and a final object $F$. Then $(\mathfrak{D}, \coprod, \times, E, F, ^*)$ is a continuous Kleene-algebra.
\end{corollary}

An important consequence of Theorem \ref{theoclosureoperator} is that the following actually defines a preorder:
\begin{definition}
Let Theorem \ref{theoclosureoperator} hold for $\mathfrak{D}(\mathcal{P}, \mathcal{S}, \times, \preceq)$. For $f, g \in \mathcal{P}$, define $f \leq_{wtt} g$, if $f \leq_m g^*$, or equivalently $f^* \leq_m g^*$ holds. The resulting degrees are denoted by $\mathfrak{D}^*$.
\end{definition}

\begin{corollary}
$\mathfrak{D}^*$ is a lattice, and a sub-meet-semilattice of $\mathfrak{D}$.
\end{corollary}

\section{Examples}
To breathe life into the generic concept of many-one reductions in a categorical setting, we shall discuss a number of applications of the framework. These examples range from merely providing a category theoretic background for known results over providing a new structure theory for previously studied objects to outlining completely new lines of investigation. The basic example have a full subcategory of $\mult$ acting as the outer category $\mathcal{P}$, but also concrete categories over $\mult$, i.e.~categories of sets equipped with additional structure, and structure-preserving multivalued functions between them, make an appearance. Finally, we also exhibit how further degree-theoretic properties do depend on the particulars of the categories involved.

\subsection{Computable many-one reductions (Type 1)}
\label{subseccomptype1}
Let $\mathcal{C}_1$ to be the subcategory of $\mult$ containing all partial computable functions $f : \subseteq \{0, 1\}^* \to \{0, 1\}^*$. As the identity $\id_{\{0, 1\}^*}$ is computable, and the composition of computable functions yields a computable function, this actually is a category. Moreover, the computable functions are closed under the formation of products and finite coproducts, and also contain all standard projections and injections, if we use the following definition:
\begin{definition}
\label{deftype1operations}
For two multi-valued functions $f, g : \subseteq \{0, 1\}^* \mto \{0, 1\}^*$, define $(f \coprod g)$, $(f \oplus g) : \subseteq \{0, 1\}^* \mto \{0, 1\}^*$ via $(f \coprod g)(0x) = 0f(x)$, $(f \coprod g)(1x) = 1g(x)$ and $(f \oplus g)(\langle x, y\rangle) = 0f(x) \cup 1g(x)$.
\end{definition}

Thus, we find $(\mult_{|\{0, 1\}^*}, \mathcal{C}_1, \times, \preceq)$ to be a moce, and study \linebreak $\mathfrak{C}_1 := \mathfrak{D}(Rel_{|\{0, 1\}^*}, \mathcal{C}_1, \times, \preceq)$. We give the special cases of the definitions and results from Subsection \ref{subsecgeneric} below:

\begin{definition}[special case of Definition \ref{defmanyonereductions}]
For two multi-valued functions $f, g : \subseteq \{0, 1\}^* \mto \{0, 1\}^*$, define $f \leq_m g$, if there are computable functions $H, K : \subseteq \{0, 1\}^* \to \{0, 1\}^*$ with $H\langle x, y\rangle \in f(x)$ whenever $y \in g(K(x))$.
\end{definition}

\begin{corollary}[of Theorem \ref{maintheolattice}]
$(\mathfrak{C}_1, \oplus, \coprod)$ is a distributive lattice.
\end{corollary}

In $\mathcal{C}_1$, there exists both an empty domain and final domains, namely the no-where defined multi-valued function $\emptyset \subset \{0, 1\}^* \times \{0, 1\}^*$ and any $\{(x, x)\} \subseteq \{0, 1\}^* \times \{0, 1\}^*$. The corresponding degrees shall be denoted by $0, 1 \in \mathfrak{C}_1$.
\begin{proposition}
\label{propcomptype1bottom}
$1$ is the least element in $\mathfrak{C}_1 \setminus \{0\}$ and contains exactly those multi-valued functions admitting a computable choice function.
\end{proposition}

We do point out that decision problems cannot be considered as a special case of multi-valued functions in the straight-forward way, as our definition of many-one reductions allows modifications of the output; in particular, the characteristic function of a set is trivially equivalent to the characteristic function of its complement. However, many results proven for many-one reductions between search problems hold - with identical proofs - also for Turing reductions with the number of oracle queries limited to $1$, which corresponds to the notion employed here.

For example, \textsc{Yates}' result regarding the existence of minimal pairs applies here as follows:
\begin{proposition}[{\cite{yates}}]
There are $\mathbf{a}, \mathbf{b} \in \mathfrak{C}_1 \setminus \{0, 1\}$ with total representatives such that for any $\mathbf{c} \leq_m (\mathbf{a} \oplus \mathbf{b})$ that has a representative $f \in \mathbf{c}$ of the type $f: \{0, 1\}^* \to \{0, 1\}$, we find $\mathbf{c} = 1$.
\end{proposition}

However, the constraint on the type of some representative is crucial, as we will demonstrate below. Instrumental is the next technical lemma:
\begin{lemma}
\label{lemmacomptype1turing}
There are Turing functionals $\Psi$, $\Phi$, such that for all total multi-valued functions $f, g : \{0, 1\}^* \mto \{0, 1\}^*$ and for any choice function $I$ of $(f \oplus g)$, either $\Psi^I$ is a choice function of $f$ or $\Phi^I$ is a choice function of $g$.
\begin{proof}
On input $x$, $\Psi$ will search for some $y$ such that $I\langle x, y\rangle = 0z$. Once this is found, it will output $z$. On input $y$, $\Phi$ will search for some $x$ such that $I\langle x, y\rangle = 1z$, and output $z$.

It is clear that if $I$ is a choice function of $(f \oplus g)$, then $\Psi^I(x) \in f(x)$ and $\Phi^I(y) \in g(y)$ whenever the former values exist. It remains to show that for any suitable $I$, either $\Phi^I(x)$ exists for all $x$ or $\Psi^I(y)$ exists for all $y$. Assume that $\Phi^I(x)$ does not exist for some $x_0$. That means the search for some $y$ with $I\langle x_0, y\rangle = 0z$ for some $z$ was unsuccessful, hence, $I\langle x_0, y\rangle = 1z_y$ for all $y$. But this means that $x_0$ always is a solution to the search performed by $\Psi^I$, hence $\Psi^I(y)$ always exists.
\end{proof}
\end{lemma}

\begin{corollary}
If $\mathbf{a}, \mathbf{b} \in \mathfrak{C}_1$ have total representatives, then $\mathbf{a} \oplus \mathbf{b} = 1$ implies $\mathbf{a} = 1$ or $\mathbf{b} = 1$.
\begin{proof}
Let $f \in \mathbf{a}$ and $g \in \mathbf{b}$ be total. If $\mathbf{a} \oplus \mathbf{b} = 0$, then $f \oplus g$ has a computable choice function $I$. By Lemma \ref{lemmacomptype1turing}, either $\Psi^I$ is a (computable!) choice function of $f$, implying $\mathbf{a} = 1$; or $\Phi^I$ is a (computable!) choice function of $g$, implying $\mathbf{b} = 1$.
\end{proof}
\end{corollary}

Following the template of Subsection \ref{subseckleene}, we can introduce the $^*$-operation and derive the corresponding results about being a closure operator and inducing a Kleene algebra.

\begin{definition}
For $f : \subseteq \{0, 1\}^* \mto \{0, 1\}^*$ define $f^* : \subseteq \{0, 1\}^* \mto \{0, 1\}^*$ \linebreak by $f^*(0^n1\langle p_1, \ldots, p_n\rangle) = 0^n1\langle f(p_1), \ldots, f(p_n)\rangle$.
\end{definition}

\begin{proposition}
$^*$ is a closure operator.
\end{proposition}

\subsection{Polynomial-time many reductions (Type 1)}
\label{subsecpolytype1}
This time we consider the category $\mathcal{P}_1$ of polynomial-time computable partial functions $f : \subseteq \{0, 1\}^* \to \{0, 1\}^*$. This is closed under the products and coproducts given by  Definition \ref{deftype1operations}, hence, we can study $\mathfrak{P}_1 = \mathfrak{D}(\mult_{|\{0, 1\}^*}, \mathcal{P}_1, \times, \preceq)$ in the usual way. It is worth noting that the same considerations apply to other usual resource-bounded reducibilities.

\begin{definition}
For two multi-valued functions $f, g : \subseteq \{0, 1\}^* \mto \{0, 1\}^*$, define $f \leq_m^p g$, if there are polynomial-time computable functions $H, K : \subseteq \{0, 1\}^* \to \{0, 1\}^*$ with $H\langle x, y\rangle \in f(x)$ whenever $y \in g(K(x))$. Let $\mathfrak{P}_1$ denote the set of $\leq_m$-degrees of multi-valued functions of this type.
\end{definition}

\begin{corollary}[of Theorem \ref{maintheolattice}]
$(\mathfrak{P}_1, \oplus, \coprod)$ is a distributive lattice.
\end{corollary}

The empty and final domains of $\mathcal{P}_1$ are exactly those of $\mathcal{C}_1$ considered in Subsection \ref{subseccomptype1}; again we shall use $0$ and $1$ to denote the respective degrees. Also, we find the following counterpart to Proposition \ref{propcomptype1bottom}

\begin{proposition}
\label{proppolytype1bottom}
$1$ is the least element in $\mathfrak{P}_1 \setminus \{0\}$ and contains exactly those multi-valued functions admitting a polynomial-time computable choice function.
\end{proposition}

As in Subsection \ref{subseccomptype1}, the many-one degrees of decision problems (Karp degrees \cite{karp}) are not a substructure of $\mathcal{P}_1$ in the natural way; however, many results proven about them also hold for polynomial-time Turing reductions with a single permitted oracle query, which do form a natural substructure.

Some results and their proofs can even be extended to include search problems; this shall be demonstrated for \textsc{Ladner}'s density result \cite[Theorem 2]{ladner}. For this, note that two notions coinciding for single-valued functions differ for multi-valued functions, namely the existence of a computable choice function and the decidability of the graph. In accordance with Proposition \ref{propcomptype1bottom}, it makes sense to call those multi-valued functions satisfying the former condition \emph{computable}. Additionally, the latter condition has the disadvantage of not being preserved downwards by many-one reductions. However, a decidable graph is the condition needed for the following theorem. Its proof closely resembles the one of \cite[Theorem 2]{ladner}, which in turn was inspired by techniques from \cite{borodin}.

\begin{theorem}
Let $\mathbf{a}, \mathbf{b} \in \mathfrak{P}_1$ admit representatives with decidable graphs and satisfy $\mathbf{b} \nleq_m^p \mathbf{a}$. Then there are $\mathbf{b}_0, \mathbf{b}_1 \in \mathfrak{P}_1$ with $\mathbf{b} = \mathbf{b}_0 \coprod \mathbf{b}_1$, $\mathbf{b}_i \nleq_m^p \mathbf{a}$ and $\mathbf{b} \nleq_m^p \mathbf{a} \coprod \mathbf{b}_i$ for both $i \in \{0, 1\}$.
\begin{proof}
Let $a \in \mathbf{a}$ and $b \in \mathbf{b}$ both have a decidable graph. The proof constructs a polynomial-time decidable set $D \subseteq \{0, 1\}^*$ such that representatives $b_0$, $b_1$ of $\mathbf{b}_0$, $\mathbf{b}_1$ fulfilling the given criteria are obtained as $b_0(x) = 0$, $b_1(x) = b(x)$ for $x \in D$, and $b_0(x) = b(x)$, $b_1(x) = 0$ otherwise. The set $D$ will have the form $D = \{x \in \{0, 1\}^* \mid |x| \in D'\}$ for some $D' \subseteq \mathbb{N}$.

This definition of $b_0$, $b_1$ already ensures the first condition to be true. For the remain ones, a priority argument is employed. Using a enumeration $R_e$ of polynomial-time many-one reductions, and the notation $R_e(f)$ for the multi-valued function arising from the application of the reduction procedure $R_e$ to $f$, we obtain the following conditions to be satisfied:
\begin{description}
\item[($P_{4e\phantom{+0}}$)] $R_e(a \coprod b_0) \nsubseteq b$
\item[($P_{4e+1}$)] $R_e(a \coprod b_1) \nsubseteq b$
\item[($P_{4e+2}$)] $R_e(a) \nsubseteq b_1$
\item[($P_{4e+3}$)] $R_e(a) \nsubseteq b_0$
\end{description}
The polynomial-time decision procedure for $D$ now works in stages, such that on input $x$ all stages $s \leq |x|$ are performed. A clock is employed to ensure that the computation for a stage $s$ does not take longer than $cs$ steps for some fixed constant $c$. In each stage the procedure searches exhaustively for a witness verifying the condition $P_{n+1}$, where $n$ is the number of the last condition for that the search was successful. After the last stage, the procedure sets $x \in D$, iff the least number of an open condition is even.

In order to do this search, knowledge about $a$, $b$ and $D$ is needed. By assumption, the graphs of $a$ and $b$ are computable. The circularity in the definition of $D$ is resolved by aborting the search in stage $s$, if any question $x \in D?$ for $|x| \geq s$ arises. For smaller inputs, the set $D$ is already fixed at this stage. Finally, if the time for a stage runs out, the search is also aborted.

It remains to prove that for every condition a witness will eventually be found. Such a witness remains valid, hence, finding a witness proves truth of the condition. On the other hand, as the time available for the search increases unboundedly, if the condition is true, eventually a witness will be found. So let $P_l$ be the condition with the least number that remains unsatisfied, and let $s$ be the first stage in which a witness for $P_l$ was sought.

\begin{description}
\item[Case $P_l = P_{4e}$] By construction, this implies $b_0(x) = 0$ for $|x| \geq s$. Hence, $b_0$ is polynomial-time computable and $\mathbf{a} \coprod \mathbf{b}_0 = \mathbf{a}$. But then $R_e(a \coprod b_0) \subseteq b$ implies $\mathbf{b} \leq_m \mathbf{a}$ in violation to the initial assumption.
\item[Case $P_l = P_{4e+1}$] By construction, this implies $b_1(x) = 0$ for $|x| \geq s$. Hence, $b_1$ is polynomial-time computable and $\mathbf{a} \coprod \mathbf{b}_1 = \mathbf{a}$. But then $R_e(a \coprod b_1) \subseteq b$ implies $\mathbf{b} \leq_m \mathbf{a}$ in violation to the initial assumption.
\item[Case $P_l = P_{4e+2}$] By construction, this implies $b_1(x) = b(x)$ for $|x| \geq s$. Hence, $\mathbf{b}_1 = \mathbf{b}$. But then $R_e(a) \subseteq b_1$ implies $\mathbf{b} \leq_m \mathbf{a}$ in violation to the initial assumption.
\item[Case $P_l = P_{4e+3}$] By construction, this implies $b_0(x) = b(x)$ for $|x| \geq s$. Hence, $\mathbf{b}_0 = \mathbf{b}$. But then $R_e(a) \subseteq b_0$ implies $\mathbf{b} \leq_m \mathbf{a}$ in violation to the initial assumption.
\end{description}
As every case of the contrary assumption leads to a contradiction, the constructed set must fulfill the desired criteria.
\end{proof}
\end{theorem}

\begin{corollary}
The degrees in $\mathcal{P}_1$ admitting decidable graphs are dense (in themselves).
\end{corollary}

A question that has received a lot of attention regarding (polynomial-time) many-one reductions between decision problems is about the existence and nature of minimal pairs. In terms of lattice theory\footnote{Which are of course not applicable to the original setting.}, this asks whether the degree $1$ is meet-irreducible, and if not, what kind of pairs can satisfy $\mathbf{a} \oplus \mathbf{b} = 1$. Following the initial result by \textsc{Ladner} that minimal pairs for polynomial-time many-one reductions between decision problems exist \cite{ladner}, \textsc{Ambos-Spies} could prove that every computable super-polynomial degree is part of a minimal pair \cite{ambosspies}.

For search problems, however, the question remains open:
\begin{question}
\label{question:1p1}
Is $1 \in \mathcal{P}_1$ meet-irreducible?
\end{question}

The techniques used to construct a minimal pair in \cite{ladner, ambosspies} diagonalize against pairs of reductions $R_e$, $R_f$ trying to prevent $R_e(a) = R_f(b)$ for the constructed representatives $a, b$. If the equality cannot be prevented, then one can prove that the resulting set is already polynomial-time decidable using a constant prefix of $b$, hence, polynomial-time decidable. However, for search problems any pair of reductions to a pair of search problems produces a search problem, namely $R_e(a) \cup R_f(b)$.

A non-computable minimal pair for Type-2 search problems was constructed in \cite{paulykojiro}. Here, the crucial part is the identifiability of hard and easy instances, which is not available in a Type-1 setting. The negative answer we obtained for computable many-one reductions in Subsection \ref{subseccomptype1} relied on Lemma \ref{lemmacomptype1turing}, which again cannot be transferred to the time-bounded case: There are polynomial-time decidable relations $R$ such that neither $R$ nor its inverse $\neg R^\dag$ admit a polynomial-time choice function, even if $P = NP$ should hold\footnote{A counterexample can be constructed as follows. On input $(x, y)$, the decision procedure works in stages $i$, starting at $i = 1$. In stage $2i$, it tests $|x| \leq i \wedge |y| \leq 2^{i}$, deciding $(x, y) \in R$ if yes, and proceeding to the next stage otherwise. In stage $2i + 1$, it tests $|x| \leq 2^{i} \wedge |y| \leq i$, deciding $(x, y) \notin R$ if yes, and proceeding to the next stage otherwise.}.

\subsection{Many-one reductions between sets}
In the two previous subsections, we warned that the traditional many-one reductions between sets are not special cases of the many-one reductions between multivalued functions discussed there. This is due to the fact that our definition permits modification of the output, in particular usually the (characteristic function of) a set will be equivalent to its complement.

Nonetheless, our framework is sufficiently powerful to include many-one reductions between sets. For this, we simply need a two-element object $2$ such that for any other object $X \in Ob(\mathcal{S})$ and any morphism $f : 2 \times X \to X$ we find $f = \pi_1^{2, X}$. In this way, sets and their complements can potentially be distinguished, and the post-processing allowed in our framework becomes useless. To give an example for such a situation: If we want to describe computable many-one reductions between sets, we can pick two Turing-incomparable sequences to represented \emph{yes} and \emph{no}\footnote{This answers a question asked by Mathieu Hoyrup at CCA 2011.}, and thus obtain a substructure of Weihrauch-reducibility discussed in the next subsection.

\subsection{Weihrauch reducibility (Type 2)}
As mentioned before, the investigation of Weihrauch reducibility served as the prototype for the present approach to many-one reductions between multivalued functions. The degree structure can be obtained by choosing $\mathcal{P}$ as the full subcategory of $\mult$ induced by the object $\Baire$, and by picking $\mathcal{S}$ as the category of partial computable functions on Baire space $\Baire$.

More faithful to the applications, the category of multivalued functions between represented spaces might serve as $\mathcal{P}$, with $\mathcal{S}$ being its subcategory of the computable multivalued functions between represented spaces. Both categories are examples of concrete categories over (subcategories of) $\mult$. For an overview on the theory of represented spaces, we refer to \cite{pauly-synthetic}.

As there is extensive literature on Weihrauch reducibility \cite{brattka2, brattka3, paulybrattka, paulybrattka2, gherardi,gherardi4, paulyincomputabilitynashequilibria, paulykojiro, paulybrattka3cie}, we shall not discuss its structure here in detail. It is worth pointing out that the r\^ole of the category extension of the continuous (multivalued) functions between represented spaces over the computable (multivalued) functions also influences the investigation of the reducibilities, as besides Weihrauch reducibility the variant of continuous Weihrauch reducibility appears. In light of the category-theoretic background, it is no surprise that both reducibilities behave very similarly.

A further variation of interest would consist in replacing computable in the definition of Weihrauch reducibility by polynomial time computable, using the recent introduction of a sufficiently general notion of polynomial time computability in \cite{kawamura}. Again, based on our framework, the basic structural theory remains the same. Moreover, as shown in \cite{pauly-kawamura,pauly-kawamura-arxiv}, the Weihrauch degrees form a substructure of the polynomial-time Weihrauch degrees. A very restricted version of type-2 polynomial time many-one reductions between multivalued function already appeared in \cite{beame}.

\subsection{Parameterized Search Problems}
\label{subsec:parameterized}
While parameterized complexity theory \cite{downeyfellows, flum} is certainly well-established, parameterized search problems are only cursorily touched upon in \cite{gottlob2}. However, the main ideas can readily be developed in our framework.

The crucial new element in parameterized complexity theory are parameterizations, which are usually defined to be polynomial-time computable functions $\kappa : \{0, 1\}^* \to \mathbb{N}$ \cite[Definition 1.1]{flum}. We shall take the broader approach of considering any function $\kappa : \{0, 1\}^* \to \mathbb{N}$ as a parameterization initially. The parameterizations are the objects in our categories $\mathcal{P}_{psp}$, $\mathcal{S}_{psp}$. A $\mathcal{P}_{psp}$-morphism from $\kappa_1$ to $\kappa_2$ is any triple $(\kappa_1, \kappa_2, R)$ where $R \subseteq \{0, 1\}^* \times \{0, 1\}^*$ is a partial multi-valued function. In particular, all hom-sets $\mathcal{P}_{psp}(\kappa_1, \kappa_2)$ are isomorphic in $\mathcal{P}_{psp}$.

The morphisms in $\mathcal{S}_{psp}$ are those $(\kappa_1, \kappa_2, R)$ where $R$ is the graph of some function $f$ such that there is an algorithm computing $f$ with run-time bounded by $t(\kappa_1(w)) \cdot p(|w|)$ where $t$ is some computable function and $p$ some polynomial, and $f$ furthermore satisfies $\kappa_2(f(w)) \leq F(\kappa_1(w))$ for some computable function $F: \mathbb{N} \to \mathbb{N}$. It is clear that $\mathcal{S}_{psp}$ is closed under suitable composition, i.e. indeed a category.

Now we need products and coproducts of parameterizations. We can define these via $(\kappa_1 \times \kappa_2)(\langle u, v\rangle) = \max \{\kappa_1(u), \kappa_2(v)\}$ and $(\kappa_1 + \kappa_2)(iu) = \kappa_i(u)$. This allows us to define products and coproducts of morphisms in $\mathcal{P}_{psp}$ by demanding $(\kappa_1, \kappa_2, R) \times (\kappa_1', \kappa_2', Q) = (\kappa_1 \times \kappa_1', \kappa_2 \times \kappa_2', R \times Q)$ and $(\kappa_1, \kappa_2, R) \coprod (\kappa_1', \kappa_2', Q) = (\kappa_1 + \kappa_1', \kappa_2 + \kappa_2', R \coprod Q)$. Straight-forward calculation verifies that these actually are products and coproducts (when amended with suitable projections and injections), and that $\mathcal{S}_{psp}$ is closed under them.

Finally, we define $(\kappa_1, \kappa_2, R) \preceq (\kappa_1, \kappa_2, Q)$ to hold, iff $\dom(R) \subseteq  \dom(Q) \wedge R \subseteq Q$ holds. We find that $\preceq$ commutes with coproducts, products and composition, and that coproducts and products also commute. The infimum $\inf \{\iota_1\pi_1, \iota_2\pi_2\}$ always exists, hence, we find the parameterized search problems to form a distributive lattice as a corollary of Theorem \ref{maintheolattice}, which we shall denote by $\mathfrak{P}_{psp}$.

At first it may seem surprising that a parameterized search problem has {\bf two} parameterizations, not only one for the input, but also for the output. This does enable some nice structural results, for example the identity as a map from the parameterization $\kappa_{\bot}$ to the parameterization $\kappa_{\top}$, where $\kappa_{\bot}(x) = 1$ and $\kappa_{\top}(x) = |x| + 1$, turns out to be complete for computable parameterized search problems.

However, for applications one might prefer to use a single parameterization to specify a search problem. In particular, this is a necessary step to consider parameterized approximation problems and parameterized counting problems as a special case of parameterized search problems. This can be achieved by fixing $\kappa_{\bot}$ as the parameterization on the output side. One readily verifies $\kappa_\bot + \kappa_\bot = \kappa_\bot \times \kappa_\bot = \kappa_\bot$, hence, this restriction is compatible with the lattice operations. The definition of reductions then takes the form:

\begin{definition}
For simply parameterized search problems $(\kappa_1, P)$, $(\kappa_2, Q)$, let $P \leq_m Q$ hold, if there are functions $F$, $G$ such that $x \mapsto F(\langle x, gG(x)\rangle)$ is a selector of $P$ for any selector $q$ of $Q$, additionally satisfying that $F(\langle x, y\rangle)$ is computable in time $f(\kappa_1(x)) \cdot p(|x|+|y|)$ and $G(x)$ is computable in time $f(\kappa_1(x)) \cdot p(|x|)$ for some computable function $f$ and some polynomial $p$, and furthermore that $\kappa_2(G(x)) \leq g(\kappa_1(g))$ for some computable function $g$.
\end{definition}

We find empty and initial domains in both $\mathcal{P}_{psp}$ and $\mathcal{S}_{psp}$, namely the no-where defined multi-valued function with arbitrary parameterization, and all multi-valued functions $x \mapsto x$ for some constant $x \in \{0, 1\}^*$, again with arbitrary parameterization. As in Subsections \ref{subseccomptype1}, \ref{subsecpolytype1} we refer to the respective degrees by $0$ and $1$. We find again a counterpart to Propositions \ref{propcomptype1bottom}, \ref{proppolytype1bottom}:

\begin{proposition}
\label{propparameterbottom}
$1$ is the least element in $\mathfrak{P}_{psp} \setminus \{0\}$ and contains exactly those parameterized multi-valued functions admitting a fixed parameter tractable choice function.
\end{proposition}

This in turn shows us that it is reasonable to consider any non-$\{0 ,1\}$ degree of parameterized search problems as intractable. Comparison with the suggestion made before \cite[Theorem 4.2]{gottlob2} to regard a parameterized search problem as intractable, iff its tractability would imply tractability of a decision problem regarded as intractable invites the following question:
\begin{question}
Is there a non-fixed parameter tractable single-valued parameterized search problem with binary image below any non-fixed parameter tractable parameterized search problem?
\end{question}

\subsection{Medvedev-reducibility}
While Medvedev-reducibility \cite{medvedev} is not commonly regarded as a many-one reduction, we can nonetheless apply Theorem \ref{maintheolattice} in order to prove that it is a distributive lattice. For this, we choose $\mathcal{P}_{\mathfrak{M}}$ to be all computable partial multi-valued functions on Baire space, and $\mathcal{S}_{\mathfrak{M}}$ to be all computable partial functions on Baire space (equivalently, we could use $\mathcal{S}_\mathfrak{M} := \mathcal{P}_\mathfrak{M}$). Proceeding as before for the remaining parts of a moce, we obtain just the {\bf dual} of Medvedev reducibility, as demonstrated in \cite{paulykojiro}.

\section{Outlook}
There are two main avenues of further research based upon the presented work: First, the specific settings of many-one reductions discussed in the preceding section contain a variety of interesting open questions. The parameterized search problems of Subsection \ref{subsec:parameterized} seem to be particularly virgin territory. Question \ref{question:1p1} also seems to have the potential of inspiring new techniques in degree-theoretic complexity theory.

Second, we do not claim to have spoken the last word on a categorical treatment of multivalued functions -- our goal rather is to demonstrate the relevance of such an approach, and to provide some preliminary results. A deeper investigation would probably be linked to looking into poset-enriched restriction categories. The result of Proposition \ref{proppreceqdomains} would offer itself as a definition there:

\begin{definition}
A poset-enriched restriction category is a category $\mathcal{C}$ with both a poset-enrichment structure $(\leq_{A,B})_{A,B \in \textrm{Ob}(\mathcal{C})}$ and a restriction endofunctor $\overline{\phantom{f}}$ satisfying $f \leq g \Rightarrow \overline{f} \leq \overline{g}$.
\end{definition}

Natural questions would be whether completing the poset-enrichment to a meet-semilattice enrichment can somehow be related to moving from $\parti$ to $\mult$; and how the inherent poset-enrichment obtainable from the restriction functor via $f \subseteq g :\Leftrightarrow f = g \circ \overline{f}$ interacts with the explicit poset-enrichment. Regarding the latter, it would be interesting to see whether Condition 4 in Definition \ref{def:multivaluedcategory} arises in a more natural way.

A general understanding of how exponentials behave in such categories might allow to generalize the development of the composition $\star$ for Weihrauch degrees \cite{paulybrattka4} to the generic setting of the present paper.

Other recent developments that should be mentioned are the observation that restriction categories can be understood in terms of enriched category theory in \cite{cockett5}; as well as an approach to understand Weihrauch reducibility in categorical terms based on fibrations in \cite{yoshimura}.

\bibliographystyle{alpha}
\bibliography{../spieltheorie}

\section*{Acknowledgements}
The author is tremendously grateful for the very helpful referee reports.

\label{lastpage}

\end{document}